\documentclass{article}
\usepackage{graphicx,color,array}
\usepackage{booktabs}
\usepackage{longtable}
\usepackage{harvard}

\title{Introduction to Quantum Algorithms for Physics and Chemistry}
\author{Man-Hong Yung$^1$, James D. Whitfield$^{2,3}$, Sergio Boixo$^{1,4}$,\\ David G. Tempel$^5$, and Alan Aspuru-Guzik$^1$}

\begin{document}


\date{\today }

\maketitle
\begin{abstract}
An enormous number of model chemistries are used in computational 
chemistry to solve or approximately solve the Schr\"odinger equation; each with 
their own drawbacks. One key limitation is that the hardware used in
computational chemistry is based on 
classical physics, and is often not well suited for simulating models
in quantum physics. In this review, we focus on applications of quantum computation to chemical physics problems. We describe the algorithms that have been proposed for the electronic-structure problem, the simulation of chemical dynamics, thermal state preparation, density functional theory and adiabatic quantum simulation. 
\end{abstract}

\begin{center}$^1$Department of Chemistry and Chemical Biology, Harvard University, Cambridge, MA\\
$^2$NEC Laboratories America, Princeton, NJ \\
$^3$Physics Department, Columbia University, New York, NY \\
$^4$Information Sciences Institute, University of Southern California, Marina del Rey, CA\\
$^5$Department of Physics, Harvard University, Cambridge, MA
\end{center}

\tableofcontents

\section{Introduction}
Controllable quantum systems provide unique opportunities for solving problems in quantum chemistry and many-body physics that are intractable by classical computers. This approach is called ``quantum simulation"\footnote{Unfortunately, the term ``quantum simulation" in the community of computational physics refers to numerical simulation of quantum systems using classical computers.}, and was pioneered by \citeasnoun{Feynman1982}. There are two different approaches for quantum simulation: analog or digital. In analog quantum simulation, dedicated physical systems are engineered to emulate the behavior of other quantum systems. A classic example is the use of atoms trapped in optical lattices to simulate the (Bose-)Hubbard model. Analog simulators are therefore special-purposed machines. On the other hand, digital simulation uses a universal quantum computer. Interestingly, a universal quantum computer is also capable, in principle, of factoring arbitrary long numbers~\cite{Shor1997a}, whereas a classical computer is not known to be able to perform the same task. For a recent review, see e.g. \citeasnoun{Ladd2010} and \citeasnoun{Hauke2011}.


One key advantage of simulations with quantum computers over classical computers is the huge Hilbert space available to faithfully represent quantum systems. Moreover, quantum simulation avoids many problems encountered in classical simulation. For example, many classical algorithms relying on Monte Carlo methods exhibit the so-called fermion sign problem that severely damages the performance of the algorithm. In quantum simulation, this problem can be avoided by either encoding the fully anti-symmetrized wavefunction in the qubit states, or by performing Jordan-Wigner transformations in the second quantized Hamiltonian and turn it into a spin Hamiltonian as first suggested by~\citeasnoun{Ortiz2001}. The latter case will often result in non-local interaction terms, but it can still be simulated efficiently in a quantum computer. 

The purpose of this article is to introduce the basic concepts of digital quantum simulation and several recent developments achieved in our group. We note that this is by no means a comprehensive review of the whole literature in quantum simulation. We will selectively cover materials that we find most useful to convey an overall idea about the current status of quantum digital simulation. Several review articles~\cite{Zalka1998b,Buluta2009a,Brown2010a,Kassal2011} and book chapters~\cite{Kitaev2002,stolze2008quantum,nielsen2011quantum,williams2010explorations} already present a different emphasis. This review also contains some new material. Sections~\ref{section:1st_quant} and \ref{sec:2nd_quant} present new descriptions of the simulation in the first and second quantized representations, respectively. Section \ref{sec:perturbative_update} lays out a new point of view for the perturbative update of thermal states from smaller to bigger quantum systems, and a new bound for the change of a thermal state due to a perturbation. 

\subsection{Quantum computational complexity and chemistry}
\subsubsection{An exponential wall for many-body problems}
The theory of computational complexity studies the scaling of the resources necessary to solve a given problem as a function of input size. Problems are considered to be ``easy," or efficiently solvable, if the time (or number of steps) for solving the problem scales as a polynomial of the input size $n$. For example, sorting a list of $n$ items will take at most $O(n^2)$ steps. On the other hand, problems are considered ``hard" if the scaling is exponential in $n$. This exponential scaling is essentially true in the worst case for almost all many-body problems in physics and chemistry~\cite{Pople1999}. A concise discussion of this point is given by~\citeasnoun{Kohn1999}, where the exponential scaling of the Hilbert space of many-electron problems is referred to as the ``Van Vleck catastrophe." The argument presented is as follows: if for each molecule, the accuracy to which one can approximate the state is $(1-\epsilon)$ (under a suitable metric), then for $n$ non-overlapping (and non-identical) molecules, the approximation worsens exponentially as $(1-\epsilon)^n$. In the next subsection, we discuss the connection of many-body problems with computational complexity further.

\subsubsection{Computational complexity of quantum simulation}
The study of the computational complexity of problems in quantum
simulation helps us better understand how quantum computers can
surpass classical computers. It has also spurred new
developments in computational complexity. For simplicity, 
computational complexity is often formulated
using decision problems. A decision problem resolves if some condition
is true or false e.g. is the ground-state energy of the system below a
certain critical value?  Although the answer to decision problems is
either ``yes" or ``no," one can keep asking questions in a binary
search fashion. For instance, one could attempt to determine in this way the ground-state energy to an arbitrarily high accuracy.

A complexity class contains a set of computational problems that share some common properties about the computational resources required for solving them. We briefly summarize below a few important examples of complexity classes of decision problems.

\subsubsection*{$\mathsf{P}$ and $\mathsf{NP}$ problems} 
The complexity class~$\mathsf{P}$ contains all decision problems that are solvable in a polynomial time with a classical computer (more precisely, a deterministic Turing machine). Roughly speaking, solving a problem in a polynomial time refers to the cases where the number of steps for solving the problem scale as a polynomial power instead of exponentially. This is considered ``efficient'' but, of course, there could be exceptions. For example, problems that scale as $O(n^{10000})$ may take very long time to finish, compared with ones that scale exponentially as $O(1.0001^n)$. 

Nevertheless, from a theoretical perspective, this division allows for considerable progress to be made without considering the minutiae of the specific system or implementation. However from a practical standpoint, the order of the polynomial may be very important; especially in chemistry where an algorithm is applied to many molecules and many geometries.  That said, the notion of polynomial versus exponential makes sense when considering Moore's ``law:'' the density of transistors in classical computers doubles every two years\footnote{The exponential growth in the computational density is expected to cease sometime this century highlighting the importance of new methods of computation such a quantum computation. The growth in CPU clock speed has already ceased.}. If the algorithm runs in exponential-time, one may be forced to wait several lifetimes in order for an instance to become soluble due to better classical hardware.  


Practically, common hard problems typically fall into the complexity class~$\mathsf{NP}$, which contains decision problems whose ``yes" instances can be efficiently verified to be true with a classical computer given an appropriate ``solution'' or witness. There is no doubt that $\mathsf{P}$ is a subclass of $\mathsf{NP}$, i.e., 
\begin{equation}
{\sf P} \subset {\sf NP} \quad. 
\end{equation}
 As an example, finding the prime factors of an integer belongs to an $\sf NP$ problem; once the factors are given, then it is easy to check the answer by performing a multiplication. Interestingly, finding the ground state energy of the Ising model
\begin{equation}
	\sum_{(i,j) \in E}  \sigma_z^i \sigma_z^j +\sum_{i \in V} \sigma_z^i\;,
\end{equation}
where $(V,E)$ is a planar graph, is an NP-complete~\cite{barahona_computational_1982}. This implies that if a polynomial algorithm for finding the ground state energy is found, then all of the problems in $\mathsf{NP}$ could be solved in polynomial time. In other words, it will imply ${\sf P} = {\sf NP}$, a result considered highly unlikely. A rigorous proof or disproof of this statement would constitute a significant breakthrough\footnote{${\sf P}$ vs ${\sf NP}$ is one of the Millennium Problems of the Clay Mathematics Institute [${\rm http://www.claymath.org/millennium/P\_vs\_NP/}$]}. 

It is believed, but not known with certainty, that quantum computers are not capable of solving all $\sf NP$ problems efficiently. Nevertheless, as mentioned above, they can solve the integer-factoring problem efficiently. It is believed that the complexity of integer-factoring is intermediate between {\sf P} and {\sf NP}~\cite{Shor1997a}. 

\subsubsection*{$\mathsf{BQP}$ and $\mathsf{QMA}$ problems} 
The quantum analog of ${\sf P}$ and $\mathsf{NP}$ problems are, respectively, the $\sf BQP$ (bounded-error quantum polynomial time) and $\sf QMA$ (quantum Merlin Arthur) problems\footnote{More precisely, $\sf BQP$ is analogous to the classical complexity class $\sf BPP$, which refers to problems that can be solved with randomized algorithms in a classical computer in polynomial time, subject to a bounded error probability.}. $\sf BQP$ is the class of (decision) problems that are solvable by a quantum computer in polynomial time. $\sf QMA$ is the class of (decision) problems that can be verified by a quantum computer in polynomial time. Like $\sf NP$-problems, the $\sf QMA$ class covers many problems that are important to physics and chemistry~\cite{Liu2007,Schuch2009,Wei2010}. For example, the ground-state problem of Hamiltonians involving local interaction terms is known to be $\sf QMA$-complete~\cite{Kitaev2002,Kempe2004}. For more discussion on topics of computational complexity and quantum simulation, readers may find the following references useful: \citeasnoun{Aharonov2002}, \citeasnoun{Rassolov2008}, \citeasnoun{Aaronson2009}, \citeasnoun{Osborne2011} and \citeasnoun{Aaronson2011a}. 

The key point here is that so far it is not known whether quantum computers can solve $\sf NP$ and $\sf QMA$ problems efficiently. In fact, many attempts (see e.g. \citeasnoun{Young2008}, \citeasnoun{Poulin2009c}, \citeasnoun{Poulin2009a}, \citeasnoun{Young2010a}, \citeasnoun{Bilgin2010}, \citeasnoun{Yung2010}, and \citeasnoun{Hen2011}) show that exponential resources are required to solve problems in these classes. Nevertheless, many problems in physics and chemistry do exhibit symmetries and structures that we could exploit to construct efficient quantum simulation algorithms. This is the main theme of the discussion in the rest of the paper.  

\subsection{Basic quantum algorithms for digital quantum simulation}
Digital quantum simulation cannot be easily understood without a detour into the basics of quantum algorithms. Quantum algorithms are procedures for applying elementary quantum logic gates to complete certain unitary transformations of the input state. The quantum computer state is usually written in terms of qubits (two-level systems). In the two-dimensional Hilbert space of a single qubit, we label the upper and lower eigenstates of $\sigma^z$ as $|0\rangle$ and $|1\rangle$. Note that the choice of $\sigma^z$ as the computational basis is arbitrary.  This is called the computational basis, and the matrix representation of operators and states are written in this basis unless otherwise stated.  The unitary transformations of the qubits may be visualized using quantum circuit diagrams introduced later to explain some of the more complex quantum algorithms.  

It is known that any unitary gate can be decomposed into some sets of universal quantum logic gates that contains single- and two-qubit operations~\cite{nielsen2011quantum}.  The first gate of interest is the single-qubit Hadamard transformation defined (in the computational basis) as
\[
{\sf H}=
\frac{1}{\sqrt{2}}
\left[
\begin{array}{cr}
	1 & 1\\
	1 & -1
\end{array}
\right]
\]
The Hadamard gate transforms between the $\sigma^z$ basis and the $\sigma^x$ basis ($|\pm\rangle=(|0\rangle\pm |1\rangle)/\sqrt{2}$) and will be used throughout the article.  A second gate of interest is the CNOT (controlled not) gate which is a non-trivial two-qubit gate. It leaves one input qubit  unchanged and acts with $\sigma^x=|0\rangle\langle 1|+|1\rangle\langle 0|$ on the second qubit when the first qubit is in the state $|1\rangle$. The first qubit is called the control and the NOT operation is applied coherently when the control qubit is in a superposition of computational basis states. Symbolically, the gate is written as CNOT$=|1\rangle\langle1|\otimes \sigma^x + |0\rangle\langle0|\otimes I$. The Hadamard and CNOT gates are not universal for quantum computation, and in fact quantum algorithms with only these gates can be simulated efficiently classically  as shown by the Knill-Gottesman theorem~\cite{nielsen2011quantum}.  Therefore, this gate set must be augmented by other single qubit gates which can always be expressed by single-qubit rotations, $R_x$, $R_y$, and $R_z$ {where} $R_x$ is defined as $\exp[-i \sigma^x \theta/2]$ for real angle $\theta$. 

There are two elementary algorithms, namely quantum Fourier transform (QFT) and phase estimation algorithm (PEA), that play important roles in many applications in quantum simulation. We turn our attention to them now.

\subsubsection{Quantum Fourier transform (QFT)}
Given a vector with $N$ elements $( {{x_0},{x_1},..,{x_{N - 1}}})$, in classical computation, the discrete Fourier transform outputs another vector of $N$ numbers $({{y_0},{y_1},..,{y_{N - 1}}})$ through the following relation:
\begin{equation}\label{QFT:FFT}
{y_k} = {1 \over {\sqrt N }}\sum\limits_{j = 0}^{N - 1} {{x_j} {e^{2\pi ijk/N}}} \quad.
\end{equation}
In quantum computation, for any given quantum state, $\left| \phi  \right\rangle  = \sum\nolimits_{x = 0}^{N - 1} {\phi \left( x \right)\left| x \right\rangle }$, the goal of the quantum Fourier transform $U_{\rm QFT}$ is to perform the following unitary transformation:
\begin{equation}\label{QFT:QFT}
{U_{\rm QFT}}\left| \phi  \right\rangle  = \sum\limits_{k = 0}^{N - 1} {\tilde \phi \left( k \right)\left| k \right\rangle } \quad ,
\end{equation}
where $\tilde \phi \left( k \right) = ( {1/\sqrt N } )\sum\nolimits_{x = 0}^{N - 1} {\phi \left( x \right)} {e^{2\pi ixk/N}}$ is the Fourier-transform of the function ${\phi \left( x \right)}$ (compare with Eq.~(\ref{QFT:FFT})). Due to the linearity of $U_{\rm QFT}$, it is sufficient to consider the transformation of the basis vectors such that 
\begin{equation}
{U_{\rm QFT}}\left| x \right\rangle  = {1 \over {\sqrt N }}\sum\limits_{k = 0}^{N - 1} {{e^{2\pi ixk/N}}\left| k \right\rangle } \quad .
\end{equation}

For a system of $n$ qubits, the number of gates required for such a transformation is $O(n^2)$~\cite{nielsen2011quantum}. For the classical case (see Eq.~(\ref{QFT:FFT})), one will require $O(n 2^n)$ gates to complete the same transformation, e.g. with Fast Fourier transform (FFT). This may seem to suggest that quantum computers are exponentially more efficient in performing the task of discrete Fourier transformation. However, the caveat is that one cannot directly compare QFT with the classical FFT. The reason is that if we want to obtain a particular Fourier-transform coefficient, say $\tilde \phi \left( k \right)$, from the quantum state in Eq.~(\ref{QFT:QFT}), it would still require exponentially many steps to extract the information (phase and amplitude), e.g. through quantum state tomography where many measurements are used to analyze the state~\cite{nielsen2011quantum}.

Nevertheless, QFT is essential in many applications in digital quantum simulation. As we shall see in section~{\ref{section:1st_quant}}, it allows us to simulate the time dynamics of particles efficiently by moving between the position and momentum representations. Another important application of the QFT is phase estimation, discussed next.

\subsubsection{Phase estimation algorithm (PEA)}
The phase estimation algorithm $U_{\rm PEA}$ is an essential component for many quantum algorithms for quantum simulation, as well as the celebrated factoring algorithm~\cite{Shor1997a}. Loosely speaking, the PEA can be considered as a realization of the von Neumann measurement scheme (without the last projective measurement) in the eigenvalue basis $\left| {{a_k}} \right\rangle$ of any Hermitian observable $A$ (e.g. Hamiltonian $H$). More precisely, if we prepare a register of $m$ ancilla qubits initialized in the state $\left| {000...0} \right\rangle $, then for any given state $\left| \phi  \right\rangle  = \sum\nolimits_k {{c_k}\left| {{a_k}} \right\rangle }$, we have 
\begin{equation}\label{PEA:projective_m}
{U_{\rm PEA}}\left| \phi  \right\rangle \left| {000...0} \right\rangle  \approx \sum\limits_k {{c_k}\left| {{a_k}} \right\rangle } \left| {{A_k}} \right\rangle \quad,
\end{equation}
where, for the moment, we assume that the $A_k$'s are the $m$-integer-digit representation (i.e., ${A_k} \in \left\{ {0,1,2,..,{2^m} {-} 1} \right\}$) of the eigenvalue of $A$. xThe projective measurement cannot be implemented perfectly in general (hence the $\approx$ symbol). We will see where the errors come from as we go through the details of the algorithm below.

Suppose that we are given an eigenstate $\left| {{a_k}} \right\rangle$ of the Hermitian observable $A$. The goal of PEA is to determine $A_k$, given that we are able to simulate a unitary operator $W$ where 
\begin{equation}\label{PEA:phase}
W\left| {{a_k}} \right\rangle  = {e^{2\pi i{\phi_k}}}\left| {{a_k}} \right\rangle \quad,
\end{equation}
and $\phi_k \equiv A_k / 2^m$. The first step of the PEA is to apply Hadamard gates to each of the ancilla qubits. This results in an equal superposition of states
\begin{equation}
\left| S \right\rangle \equiv {1 \over {\sqrt {{2^m}} }}\sum\limits_{x = 0}^{{2^m} {-} 1} {\left| x \right\rangle } 
\end{equation}
where $x$ is a $m$-digit binary number. Then, taking each ancilla qubit $j$ as a control qubit, we apply the controlled-$W^{2^j-1}$ gate to the state $\left| S \right\rangle \left| {{a_k}} \right \rangle  $; this effectively 
performs the following operation:
\begin{equation}
\left| x \right\rangle \left| {{a_k}} \right\rangle  \to \left| x \right\rangle {W^{x}}\left| {{a_k}} \right\rangle \quad.
\end{equation}
Of course, from Eq.~(\ref{PEA:phase}), the right-hand side gives only a phase factor, namely $\exp \left( {2\pi ix{\phi_k}}\right)$. The resulting state is 
\begin{equation}\label{PEA:PEA}
\left( {{1 \over {\sqrt {{2^m}} }}\sum\limits_{x = 0}^{{2^m} - 1} {{e^{2\pi ix{\phi_k}}}\left| x \right\rangle } } \right)\left| {{a_k}} \right\rangle \quad .
\end{equation}
Comparing this state with that in Eq. (\ref{QFT:QFT}), and assuming the special cases where the phase angle $\phi_x$ can be expressed exactly by $m$ binary digits, the application of the inverse of the quantum Fourier transform $U_{\rm QFT}$ will convert the state in Eq.~(\ref{PEA:PEA}) into the following state
\begin{equation}
\left| {{A_k}} \right\rangle \left| {{a_k}} \right\rangle \quad.
\end{equation}
Since the unitary operator $U_{\rm PEA}$ is linear, the procedure applies to any initial state. For this particular case, where $A_k$'s are integers, we have shown that PEA is effectively a projective measurement as advertised in Eq.~(\ref{PEA:projective_m}). 

For the general case, where the $A_k$'s are real numbers, the corresponding $\phi_k$'s will have precision beyond $1/2^m$; this is the source of the errors in the expression of Eq.~(\ref{PEA:projective_m}). The overall error decreases when we increase the number of ancilla qubits and perform several QFTs in parallel (we refer to \citeasnoun{kaye_introduction_2007} for a detailed error analysis). More precisely, if we want to achieve a $p$-bit precision of $\phi_k$ with an error less than $\epsilon$, one will need more than $m = p + \log \left( {2 + 1/2\epsilon } \right)$ ancilla qubits. In general, implementing the operator $W^k$ requires $k$ times as many resources as those needed for simulating $W$. Therefore, the scaling of the quantum gates of PEA grows exponentially when we increase the precision $p$ of the phase measurement. This result is consistent with that of the general sampling theory in classical signal processing, where the precision of the Fourier spectrum $\delta \omega$ goes as the inverse of the total time $T$ sampled, i.e., $\delta \omega \sim O(1/T)$. This is because the cost of the quantum simulation is proportional to $T$, and $T$ grows exponentially with the number of bits of precision.



\section{Digital quantum simulation}
\subsection{Overview}
Broadly speaking, the steps involved in carrying out a digital quantum simulation consist of three parts: state preparation, time evolution, and measurement of observables. Measurement of Hermitian observables can be achieved via the phase estimation method~\cite{Abrams1999,Jaksch2003,knill_optimal_2007} described before. Other applications~\cite{Lidar1999,Wu2002,Somma2002,Somma2003,Byrnes2006a,Kassal2009} or quantities of physical interest such as the partition function~\cite{Master2003,Wocjan2009a}, can be obtained through variants of the methods employed in state preparation and time evolution, and we will skip them in this review. Below we give an overview of state preparation and simulation of time evolution. It turns out that many methods of state preparation also depend on the time evolution itself. Therefore, we will first cover the methods of time evolution before state preparation.

\subsection{Simulation of time evolution}\label{sec:simulation_of_time_evolution}
The simulation of the time evolution of quantum state $\left| \psi  \right\rangle$ under Hamiltonian $H$ according to the Schr\"odinger's equation ($\hbar=1$),
\begin{equation}
i \frac{\partial }{{\partial t}}\left| \psi  \right\rangle  = H\left( t \right)\left| \psi  \right\rangle \quad ,
\end{equation}
is one of the key applications of quantum computation. If, for example, the time-evolution operator 
\begin{equation}
U(t) = \exp(-iHt)
\end{equation}
can be simulated efficiently, then the eigenvalues of $H$ can be obtained through the phase estimation algorithm\footnote{Moreover, it can also be exploited for quantum cooling (see section~\ref{Qcooling}).}. As mentioned in the introduction, \citeasnoun{Feynman1982} investigated the possibility of simulating quantum systems using another quantum system, and conjectured that there existed a class of universal quantum simulators that evolved under a Hamiltonian with local interactions. This conjecture was justified by~\citeasnoun{Lloyd1996}, who argued that any Hamiltonian 
\begin{equation}
H = \sum\limits_{i = 1}^m {H_i }
\end{equation}
which can be decomposed into $m$ local terms $\{H_i\}$ can be simulated efficiently by a universal quantum computer. Each $H_i$ term acts on at most $k$ qubits (or quantum subsystems). The key idea is based on the Trotter splitting or ``trotterization'' of all non-commuting operators,
\begin{equation}\label{time:trotter}
e^{ - iHt}  \approx \left( {e^{ - iH_1 t/n} e^{ - iH_2 t/n} ...e^{ - iH_m t/n} } \right)^n \quad,
\end{equation}
where the approximation can be made arbitrarily tight by refining the time-slicing, i.e., increasing $n$. 

There exist higher order approximations (Suzuki-Trotter formulas) which reduce the error even further. 
For instance, the second-order approximation is given by
\begin{eqnarray}
    e^{-iHt} &\approx& \left(\left(e^{-ih_1\frac{\Delta t}{2}}\cdots e^{-ih_{N-1}\frac{\Delta t}{2}}\right)e^{-ih_N\Delta t}\left(e^{-ih_{N-1}\frac{\Delta t}{2}}\cdots e^{-ih_1\frac{\Delta t}{2}}\right)\right)^\frac{t}{\Delta t} \nonumber \\ &+&O(t(\Delta t)^2)
    \label{eq:Trotter2ndOrder}
\end{eqnarray}
A quantum circuit on $n$ qubits which approximates $U(\tau)$, with
error at most $\epsilon$, is efficient if the number of one- and
two-qubit gates involved is polynomial in the scaling of the problem,
i.e., ${\rm{poly}}( {n,\tau,1/\epsilon } )$ with $\tau=t/||H||$.

\subsubsection{Suzuki-Trotter formulas}
We now briefly review the use of Suzuki-Trotter formulas in quantum simulation for time-independent sparse Hamiltonians, providing an introduction to the quantum simulation literature.  Continuing the work of \citeasnoun{Lloyd1996}, works by \citeasnoun{Aharonov2003} and \citeasnoun{Childs2004} show that black-box sparse Hamiltonians are too efficiently simulatable.  Sparsity here means that the number of elements per row is bounded by some polynomial of $n$, while the dimension of the Hilbert space is  $D=2^n$. It is also required that each matrix element can be retrieved efficiently. Ref.~\cite{Aharonov2003} used a coloring scheme to decompose the Hamiltonian into a sum of $2\times2$ block diagonal matrices.  This coloring scheme has been updated in several references~\cite{Berry2007,Berry2009,Childs2011}.  The coloring scheme and blackbox simulation will not be discussed further.

\citeasnoun{Berry2007} were the first to approach the general problem of
simulating non-commuting Hamiltonians by using higher order
Suzuki-Trotter formulas.  \citeasnoun{Papageorgiou2010} returned to
this issue and their contributions will be discussed later.  The
important results of \citeasnoun{Berry2007} are
\begin{enumerate}
\item the use of higher order Trotter-Suzuki decompositions to bound
  the number of non-commuting exponentials, $N_{exp}$, necessary to
  carry out a simulation for some amount of time $t$,
\item a proof of a
  no-go theorem for sub-linear black-box simulation and
\item 
  improvements upon the coloring scheme of \cite{Aharonov2003} and
  \cite{Childs2004} for black box simulation of sparse Hamiltonians.
\end{enumerate}
The simulations in this review
are concerned with the first two results and they will be explained in
more detail after describing the Suzuki-Trotter formulas.

M. Suzuki has studied and extended the Trotter formula essentially
continuously since 1990 and this work was reviewed in
\citeasnoun{Hatano2005}. The recursive formulas introduced by Suzuki define
a fractal pattern where a combination of forward and backward
propagation leads to an improved approximation of the desired
exponential. Suzuki defines higher order
Trotter formulas in a recursive way. Beginning with the split
operator formula, $e^{Ax/2}e^{Bx}e^{Ax/2}$, for $m$ operators, the following 
series of equations were derived:
\begin{eqnarray}
	S_2(x)&=& \left(\prod_{k=1}^m e^{{h}_kx}\right)\left(\prod_{k=m}^1 e^{{h}_kx}\right)\\
	S_4(x)&=& S_2(z_2x)^2S_2( (1-4z_2)x)S_2(z_2x)^2\\
	\vdots& &\vdots\nonumber\\
	S_{2k}(x)&=& S_k(z_kx)^2S_k( (1-4z_k)x)S_k(z_kx)^2
\end{eqnarray}
The values of the constants $\{z_j\}$ are selected so that $S_{2j}$ is correct through $2j^{th}$ order and it can be shown~\cite{Hatano2005} that $z_i=(4-4^{1/(2i-1)})^{-1}$.  If there are $m$ non-commuting Hamiltonians, then the first order approximation takes  $m=N_{exp}$ and for the split operator formula, $S_2$, the number of exponentials is $2m-1$.  In general, $2(m-1)5^{k-1}+1$ exponentials are used for the $S_{2k}$ approximant. 

For the $k^{th}$ order Suzuki-Trotter, with $m$ Hamiltonians in the
sum, and error tolerance given by $\varepsilon$, \citeasnoun{Berry2007}
gives a bound on the number of exponentials by bounding each order of
the Suzuki-Trotter formula. \citeasnoun{Papageorgiou2010} presented an
improvement by noting that the relative ratio of Hamiltonian norms is
also important. The main idea is that if some of the Hamiltonians have very
small weight, then their exponentials can be effectively ignored.

The optimal order of Trotter decomposition, $k^*$, is determined by selecting the best compromise between time-step length and a decomposition using more exponentials.  In \citeasnoun{Berry2007} this was worked out for unstructured sparse Hamiltonians $N_{exp}\geq ||H||t$.  The lower bound on the generic cost of simulating an evolution was by contradiction, and relied on the lower bounds to quantum mechanical problems based on the polynomial method \cite{Beals1998}.  This bound could be violated given sub-linear simulation time.  In a departure from the methods discussed so far, \citeasnoun{Childs2009} used a quantum walk based approach to push the scaling closer to linear in the reweighed time and \citeasnoun{Raesi2011} looked at designing quantum circuits for quantum simulation.

For problems in chemistry, it is more natural to represent Hamiltonians in terms of first- and second-quantized forms. In the following, we will describe how to exploit the special structure of molecular Hamiltonians to simulate the time dynamics in quantum computers. 


\subsubsection{First-quantized representation}\label{section:1st_quant}
In the first-quantized form, the non-relativistic molecular Hamiltonians $H$ decomposes in a kinetic $T$ and potential $V$ terms, i.e., 
\begin{equation}\label{time:Ham}
H=T+V \quad .  
\end{equation}
The kinetic term includes the contribution from the nuclei and electrons separately,  
\begin{equation}
T =  - \sum\limits_i {\frac{{\hbar ^2 }}{{2M_i }}\nabla _i^2 }  - \sum\limits_j {\frac{{\hbar ^2 }}{{2m_e }}\nabla _j^2 } \quad,
\end{equation}
where $M_i$ is the mass of the nucleus $i$, and $m_e$ is the electron mass. The potential energy term comprises of the Coulomb interaction among the nuclei, among the electrons, and between the nuclei and electrons. Explicitly:
\begin{equation}
V\left( {\bf r,R} \right) = \frac{{e^2 }}{{4\pi \varepsilon _0 }}\sum\limits_{i < j} {\frac{{Z_i Z_j }}{{\left| {{\bf R}_i  - {\bf R}_j } \right|}}}  + \frac{{e^2 }}{{4\pi \varepsilon _0 }}\sum\limits_{i < j} {\frac{1}{{\left| {{\bf r}_i  - {\bf r}_j } \right|}}}  - \frac{{e^2 }}{{4\pi \varepsilon _0 }}\sum\limits_{i,j} {\frac{{Z_i }}{{\left| {{\bf R}_i  - {\bf r}_j } \right|}}} \, ,
\end{equation}
where $e$ is the electric charge, and $Z_i$ is the charge of nuclei $i$. The coordinates of nuclei $i$ and electron $j$ are denoted by ${\bf R}_i$ and ${\bf r}_j$.We will use the notation ${\bf r} = \left( {{\bf r}_1 ,{\bf r}_2 ,{\bf r}_3 ...} \right)$ (and similarly for $\bf R$). We also ignore the spin degrees of freedom, which can be incorporated easily. 

The general wavefunction can be represented in the position basis as 
\begin{equation}
\left| \Psi  \right\rangle  = \sum\limits_{\bf r,R} {\Psi \left( {\bf r,R} \right)\left| {{\bf r}_1 {\bf r}_2 {\bf r}_3 ...} \right\rangle | {{\bf R}_1 {\bf R}_2 {\bf R}_3 ...} \rangle} \quad, 
\end{equation}
where each electronic or nuclear coordinate is represented on its own grid over $m$ qubits resulting in a total of $Bm$ qubits to represent the state of $B$ particles.
Note that the grid encoded in $m$ qubits has $2^m$ points.
The complex wavefunction $\Psi \left( {\bf r,R} \right)$ in addition to being properly normalized must also be anti-symmetrized (or symmetrized for Bosons). \citeasnoun{Abrams1997} and \citeasnoun{Ward2009a} consider the necessary anti-symmetrization process for fermions in first quantization.

To simulate the dynamics~\cite{Zalka1998b,Wiesner1996,Kassal2008}, we note that although the kinetic and potential terms do not commute with each other, both can be represented as diagonal operators in momentum and position basis respectively. By using the quantum Fourier transform $U_{\rm QFT}$, it is natural to decompose the time evolution as 
\begin{equation}
e^{ - iHt}  \approx \left( {U_{\rm QFT}^\dagger e^{ - iTt/n}U_{\rm QFT} e^{ - iVt/n} } \right)^n \quad.
\end{equation}
In fact, this method is known as the split-operator method~\cite{Feit1982,Kosloff1988}. Higher-order Suzuki-Trotter formulas can also be applied, as described before. This method was applied to quantum computing in a number of works~\cite{Wiesner1996,Zalka1998c,Zalka1998b,Strini2002,Benenti2008,Kassal2008}.  

In the context of quantum computing, it remains to find a method to induce a coordinate-dependent phase factor such that
\begin{equation}\label{time:phase_induce}
\left| {{\bf r}_1 {\bf r}_2 {\bf r}_3 ...} \right\rangle \left| {{\bf R}_1 {\bf R}_2 {\bf R}_3 ...} \right\rangle \quad  \to \quad e^{ - iV\left( {\bf r,R} \right) \delta t } \left| {{\bf r}_1 {\bf r}_2 {\bf r}_3 ...} \right\rangle \left| {{\bf R}_1 {\bf R}_2 {\bf R}_3 ...} \right\rangle \quad, 
\end{equation}
where $\delta t \equiv t/n$, and similarly for the kinetic term in the Fourier basis. An efficient method\footnote{An alternative method was proposed by \citeasnoun{Benenti2008}, but it scales exponentially with the number qubits.} is implicitly described in the book \citeasnoun{Kitaev2002} (pages 131-135), which was further developed and adapted to the chemistry context by \citeasnoun{Kassal2008}. We sketch the idea here for completeness. First, we will assume that the potential energy term is rescaled to become dimensionless, and projected into a range of integer values such that $0 \le V\left( {\bf r,R} \right) \le 2^m  - 1$, where $m$ should be sufficiently large to allow appropriate resolution of $V\left( {\bf r,R} \right)$ in the integer representation. Next, we define a more compact notation $
\left| {\bf r,R} \right\rangle  \equiv \left| {{\bf r}_1 {\bf r}_2 {\bf r}_3 ...} \right\rangle \left| {{\bf R}_1 {\bf R}_2 {\bf R}_3 ...} \right\rangle$, and an algorithmic operation $\mathcal A$ to be performed in the position basis:
\begin{equation}\label{time:algorithmic}
{\mathcal A} \left| {\bf r,R} \right\rangle \left| \bf s \right\rangle  \quad \to \quad \left| {\bf r,R} \right\rangle \left| {{\bf s} \oplus V\left( {\bf r,R} \right)} \right\rangle \quad,
\end{equation}
where $\left| \bf s \right\rangle$, $\bf s =1,2,3,...$, is a quantum state of $m$ ancilla qubits, and $\oplus$ is addition modulo $2^m$. Suppose now that the ancilla qubits are initialized in the following state: 
\begin{equation}
\left| {\bf q} \right\rangle  \equiv \frac{1}{{\sqrt M }}\sum\limits_{{\bf s} = 0}^{M - 1} {e^{2\pi i {\bf s} /M} } \left| {\bf s} \right\rangle \quad,
\end{equation}
where $M \equiv 2^m$. This state is the Fourier transform of $| 1 \rangle$. Then the desired phase generating operation described in Eq.~(\ref{time:phase_induce}) can be achieved using controlled $\sigma^z$ rotations after applying $\mathcal A$ to the state $\left| {\bf r,R} \right\rangle \left| \bf q \right\rangle$. A similar procedure is applied to the kinetic term to complete the Trotter cycle. 

An alternative approach to implement the controlled-phase operation described in Eq.~(\ref{time:phase_induce}) is the following: first include a register of qubits initialized as  $\left| \bf 0 \right\rangle $. Then in a similar (but not identical) way as that described in Eq.~(\ref{time:algorithmic}), we define the operation
\begin{equation}\label{time:0_to_V}
	{\tilde{\mathcal  A}} \left| {\bf r,R} \right\rangle \left| \bf 0 \right\rangle  \quad \to \quad \left| {\bf r,R} \right\rangle \left| { V\left( {\bf r,R} \right) \delta t} \right\rangle \quad,
\end{equation}
where we used ${{\bf 0} \oplus V\left( {\bf r,R} \right) \delta t} =
V\left( {\bf r,R} \right) \delta t $. The state $ \left| { V\left(
      {\bf r,R} \right) \delta t} \right\rangle$ is the binary
representation $\left\{ {{x_1}{x_2}{x_3}...{x_m}} \right\}$ defined
through the following equality,
\begin{equation}
V\left( {\bf r,R} \right)\delta t \equiv 2\pi \times {0.{x_1}{x_2}{x_3}...{x_m}} = 2\pi \sum\limits_{k = 1}^m {{{{x_k}} \over {{2^k}}}} \quad.
\end{equation}
Now, we can decompose the overall phase as follows,
\begin{equation}
{e^{ - iV\left( {{\bf{r}},{\bf{R}}} \right)\delta t}} = {e^{ - i2\pi {x_1}/2}}{e^{ - i2\pi {x_2}/{2^2}}}...{e^{ - i2\pi {x_m}/{2^m}}}
\end{equation}
This decomposition can be achieved through the application of $m$ local phase gates ${R_k} \equiv \left| 0 \right\rangle \left\langle 0 \right| + \exp \left( { - 2\pi i/{2^k}} \right)\left| 1 \right\rangle \left\langle 1 \right|$ for each ancilla qubit. This approach requires the ancilla to be un-computed  (i.e. the inverse of the operation in Eq.~(\ref{time:0_to_V})) in the last step.


\subsubsection{Second-quantized representation}\label{sec:2nd_quant}
The first-quantization method is universally applicable to any molecule. The shortcoming is that it does not take into account the physical symmetrization properties of the underlying quantum system.  When a suitable set of basis functions is employed, the size of the problem can be significantly reduced. This is known as the second-quantization approach in quantum chemistry, which can be extended for quantum simulation.

Most studies on quantum simulation based on first quantization methods use grids to represent wave functions, while works employing second quantization methods generally use atomic or molecular orbitals as a basis set for the wave functions. We will take the later approach here. Nevertheless, the choice of basis is not the key difference between the first and second quantization. Indeed, a basis set of delta functions (or approximations to delta functions) could be used to represent a grid within second quantization. On the other hand, the storage of the same wave function is very different in second and first quantization. For example, a two-particle wave function with the first particle at site $i$ and the second at site $j$,  is represented as $|{\rm coord}_i\rangle | {\rm coord}_j\rangle$ in first quantization, and as $|0\cdots1_i\cdots1_j\cdots00\rangle$ in second quantization.

The starting point of the second-quantization approach~\cite{Aspuru-Guzik2005c,Wang2008,Whitfield2010} is the Born-Oppenheimer approximation, where the nuclear coordinates $\bf R$ are taken to be classical variables. This allows us to focus on the electronic structure problem. Ignoring the nuclear kinetic and the nuclear-nuclear interaction terms, the molecular Hamiltonian in Eq.~(\ref{time:Ham}) can be expressed as  
\begin{equation}
H = \sum\limits_{pq} {h_{pq} a_p^\dagger  a_q }  + \frac{1}{2}\sum\limits_{pqrs} {h_{pqrs} a_p^\dagger  a_q^\dagger  a_r a_s } \quad.
\end{equation}
where the fermionic creation operator $a^\dagger_p$ creates an electron in the $p$ mode from the vacuum, i.e., $a_p^\dagger \left| {\rm vac} \right\rangle  = \left| p \right\rangle$. Denote $\chi_p({\bf r})$ as the single-particle wavefunction corresponding to mode $p$.\footnote{Here $\bf r $ refers to the coordinates of one particular electron.} Then, the explicit form for the single-electron integrals is given by 
\begin{equation}
h_{pq}  \equiv  - \int {d {\bf r}} \chi _p^* \left( {\bf r} \right)\left( {\frac{{\hbar ^2 }}{{2m_e }}\nabla ^2  + \frac{{e^2 }}{{4\pi \varepsilon _0 }}\sum\limits_i {\frac{{Z_i }}{{\left| {{\bf R}_i  - {\bf r}} \right|}}} } \right)\chi _q \left( {\bf r} \right) \quad,
\end{equation}
and the electron-electron Coulomb interaction term is,
\begin{equation}
h_{pqrs}  \equiv \frac{{e^2 }}{{4\pi \varepsilon _0 }}\int {d{\bf r}_1 d{\bf r}_2 } \frac{{\chi _p^* \left( {{\bf r}_1 } \right)\chi _q^* \left( {{\bf r}_2 } \right)\chi _r \left( {{\bf r}_2 } \right)\chi _s \left( {{\bf r}_1 } \right)}}{{\left| {{\bf r}_1  - {\bf r}_2 } \right|}} \quad.
\end{equation}
These integrals have to be pre-calculated with classical computers before encoding them into the quantum algorithms. If we keep $k$ single-particle orbitals, then there are $O(k^4)$ terms. More details of the formalism of second-quantized electronic structure theory in the Born-Oppenheimer approximation can be found in \citeasnoun{Helgaker00}.

To simulate time dynamics in a quantum computer, we can apply the same  Trotterization idea described above (see Eq.~(\ref{time:trotter})), and simulate separately the terms
\begin{equation}\label{time:2Us}
\exp({ - ih_{pq} a_p^ \dagger  a_q \delta t}) \quad {\rm and} \quad \exp({ - ih_{pqrs} a_p^\dagger  a_q^\dagger  a_r a_s \delta t}) \quad.
\end{equation}
Since the simulation of every single exponential term in a quantum computer is costly, due to error-correction overheads as discussed in \citeasnoun{Clark2009}, one simplification we can make is to group the terms of single-particle terms into two-particle terms. This is possible for electronic problems with a fixed number $N$ of electrons. Consider any $N$-fermionic state, then the identity operator $I_N$ is equivalent to a summation of the following single-body number operators,
\begin{equation}
\left( {1/N} \right)\sum\limits_s {a_s^\dagger  a_s }  \quad \Leftrightarrow \quad I_N \quad,
\end{equation}
which means that we can write
\begin{equation}
a_p^\dagger  a_q  = \frac{1}{{N - 1}}\sum\limits_s {a_p^\dagger  a_s^ \dagger  a_s } a_q \quad.
\end{equation}
The last equation is a sum of two-electron terms, and can be absorbed into the pre-computed values of $h_{pqrs}$. Now, denoting the new values as ${\tilde h_{pqrs} }$, the Hamiltonian $H$ reduces to 
\begin{equation}\label{time:H_tlided}
H = \frac{1}{2}\sum\limits_{pqrs} {\tilde h_{pqrs} a_p^\dagger  a_q^ \dagger  a_r a_s } \quad.
\end{equation}
Therefore, we are left only with simulating the two-body term in Eq.~(\ref{time:2Us}). 

One challenge we need to overcome is the fermionic nature of the operators $a^\dagger_p$ and $a_q$, which comes from the anti-symmetrization requirement of fermionic wavefunctions. A first step to overcome this challenge is to map the occupation representation to the qubit configuration. Explicitly, for each fermionic mode $j$, we represent the qubit state $\left| 0 \right\rangle _j \equiv \left| \downarrow \right\rangle_j $ as an unoccupied state, and similarly $\left| 1 \right\rangle _j \equiv \left| \uparrow \right\rangle_j  $ as an occupied state. To enforce the exchange symmetry, we apply the Jordan-Wigner transformation~\cite{Ortiz2001,Whitfield2010}: 
\begin{equation}\label{time:Jordan-Wigner}
a_j^\dagger  = \left( {\prod\limits_{m <j} {\sigma _m^z} } \right)\sigma _j^ - \quad {\rm and} \quad {a_j} = \left( {\prod\limits_{m < j} {\sigma _m^z} } \right)\sigma _j^+ \quad,
\end{equation}
where 
\begin{equation}\label{time:pm=x+y}
{\sigma ^ \pm } \equiv ({\sigma ^x} \pm i{\sigma ^y})/2 \quad.
\end{equation}
By using Eq. (\ref{time:Jordan-Wigner}) and (\ref{time:pm=x+y}), we can now write the fermionic Hamiltonian in Eq.~(\ref{time:H_tlided}) as a spin Hamiltonian involving products of Pauli matrices $\left\{ {{\sigma ^x},{\sigma ^y},{\sigma ^z}} \right\}$:
\begin{equation}
{H_{spin}} = \sum\limits_{pqrs} {\sum\limits_{abcd} {g_{pqrs}^{abcd}{\theta _{pqrs}}\sigma _p^a\sigma _q^b\sigma _r^c\sigma _s^d} } \quad,
\end{equation}
where the set of indices $\{ p,q,r,s \}$ is summed over the fermionic modes, and $\{ a,b,c,d \}$ is either $x$ or $y$. The operator ${{\theta _{pqrs}}}$ keeps track of the $\sigma_z$'s; for example, if $p>q>r>s$, we then have 
\begin{equation}
{\theta _{pqrs}} = \left( {\prod\limits_{p >i> q} {\sigma _i^z} } \right) \times \left( {\prod\limits_{r >j> s} {\sigma _j^z} } \right) \quad.
\end{equation}
The punchline here is that the Hamiltonian becomes a polynomial sum of products of spin operators, and each operator is locally equivalent to $\sigma_z$. Therefore, the non-trivial part of simulating the time dynamics of the fermionic Hamiltonian is to simulate the non-local interaction terms of the following form:
\begin{equation}
{\exp({ - ig{\sigma ^z}{\sigma ^z}{\sigma ^z}....{\sigma ^z}\delta t})} \quad,
\end{equation}
where $g$ is some constant. This can be achieved by a series of controlled-NOT together with a local operation (see e.g. Figure 4.19 of \citeasnoun{nielsen2011quantum}), or the phase generating method similar to the one described in the previous section (cf. Eq.~(\ref{time:phase_induce})). The explicit circuits for simulating the time evolution operators can be found in \citeasnoun{Whitfield2010}. 




\subsubsection{Open-system dynamics}
In quantum mechanics, the time evolution dynamics of a closed system is always described by a unitary transformation of states, $U\left( t \right)\rho\, U^\dagger  \left( t \right)$. However, non-unitary dynamics occurs when the dynamics of the system of interest $S$ is coupled to the environment $B$, as in, 
\begin{equation}
\rho _S \left( t \right) \equiv {\rm Tr}_B \left[ {U\left( t \right)\rho_{SB}\, U^\dagger  \left( t \right)} \right]\;.
\end{equation}
After some approximations this evolution can often be described by a
(Markovian) quantum master equation in Lindblad form \cite{Breuer02,Lindblad76,Gorini76},
\begin{equation}\label{OSD:master_equ}
\frac{d}{{dt}}\rho _s \left( t \right) =  - i\left[ {H_s ,\rho _s } \right] + \sum_{\alpha,\beta}m_{\alpha \beta } \left( {\left[ {\Lambda _\alpha  \rho _s ,\Lambda _\beta ^ \dagger  } \right] + \left[ {\Lambda _\alpha  ,\rho _s \Lambda _\beta ^ \dagger  } \right]} \right) \quad,
\end{equation}
where $H_s$ is the system Hamiltonian,  $m_{\alpha \beta}$ is a positive matrix, and $\Lambda_{\alpha}$ is a linear basis of traceless operators. This quantum master equation is relevant in many physical, chemical, and biological processes at finite temperature~\cite{Mohseni08,Rebentrost09}. Further, this equation has many applications in quantum information processing, including preparing entangled states (from arbitrary initial states)~\cite{Kraus2008,Krauter2011,Muschik2011,Cho2011,Muller2011}, quantum memories~\cite{Pastawski2011}, and dissipative quantum computation~\cite{Verstraete2009}. It has been shown that the quantum master equation can be simulated by a unitary quantum circuit with polynomial resource scaling~\cite{Bacon2001,Kliesch2011}. The basic idea is as follows: we first re-write the master equation (Eq. (\ref{OSD:master_equ})) in the form,
\begin{equation}
\frac{d}{{dt}}\rho _s \left( t \right) = {\mathcal L} \left( {\rho _s } \right) \quad,
\end{equation}
where $\mathcal{L}$ is a super-operator. Similar to the unitary dynamics, we can define the super-operator version of the propagator ${\mathcal K} \left( t_1, t_0 \right)$ through the relation, 
\begin{equation}
\rho _s \left( t_1 \right) = {\mathcal K} \left( t_1,t_0 \right)\left( {\rho _s \left( t_0 \right)} \right)
\end{equation}
for all values of time $t_1\ge t_0$. Suppose we consider a finite time interval $T$, which can be divided into $m$ small time intervals $\Delta t$, i.e., $T = m \Delta t$. Then similar arguments~\cite{Kliesch2011} based on Trotterization show that the following approximation,
\begin{equation}
{\mathcal K}\left( T \right) \approx \underbrace {{\mathcal K}\left( {\Delta t} \right) {\mathcal K}\left( {\Delta t} \right) {\mathcal K}\left( {\Delta t} \right)...{\mathcal K}\left( {\Delta t} \right)}_{{ m \rm{ \, times}}} \quad,
\end{equation}
indeed converges when the division size goes to zero, i.e., $\Delta t \to 0$. The remaining part of the argument is to show that each of the small-time propagator terms $\mathcal{K} (\Delta t)$ can be simulated efficiently with a quantum circuit. This is generally true if the superoperator $\mathcal{L}$ is a finite (polynomial) sum of local terms~\cite{Bacon2001}.

\subsection{State preparation}\label{sec:state_preparation}
We have discussed how quantum dynamics can be simulated efficiently
with a quantum computer, but we have not yet discussed how quantum
states of physical or chemical interest can be initialized on 
the quantum computer. In fact, both thermal and ground states of physical
Hamiltonians can be prepared by incorporating the methods of simulating the 
time dynamics, as we shall explain later in this section.

We first consider a strategy to prepare quantum states that can be
efficiently described by some integrable general function, e.g., a
Gaussian wave packet.  Before we provide a general description, it may
be instructive to consider the case of creating a general (normalized)
two-qubit state,
\begin{equation}\label{state:2qubit}
{f_{00}}\left| {00} \right\rangle  + {f_{01}}\left| {01} \right\rangle  + {f_{10}}\left| {10} \right\rangle  + {f_{11}}\left| {11} \right\rangle 
\end{equation}
from the initial state $\left| {00} \right\rangle $. First of all, we
will assume that all the coefficients $f_{ij}$'s are real numbers, as the phases can be generated by the method described in Eq.~(\ref{time:phase_induce}). Now, we can write the state in Eq.~(\ref{state:2qubit}) as 
\begin{equation}\label{state:2qubit_2}
{g_0}\left| 0 \right\rangle  \otimes \left( {{\textstyle{{{f_{00}}} \over {{g_0}}}}\left| 0 \right\rangle  + {\textstyle{{{f_{01}}} \over {{g_0}}}}\left| 1 \right\rangle } \right) + {g_1}\left| 1 \right\rangle  \otimes \left( {{\textstyle{{{f_{10}}} \over {{g_1}}}}\left| 0 \right\rangle  + {\textstyle{{{f_{11}}} \over {{g_1}}}}\left| 1 \right\rangle } \right) \quad,
\end{equation}
where ${g_0} \equiv \sqrt {f_{00}^2 + f_{01}^2} $ is the probability
to find the first qubit in the state $\left| 0 \right\rangle$, and
similarly for ${g_1} \equiv \sqrt {f_{10}^2 + f_{11}^2}$. The form in
Eq.~(\ref{state:2qubit_2}) suggests that we can use the following
method to generate the general state of Eq. (\ref{state:2qubit}) from  $\left| {00} \right\rangle $. 
\begin{enumerate}
\item Apply a rotation, such that $\left| 0 \right\rangle  \to {g_0}\left| 0 \right\rangle  + {g_1}\left| 1 \right\rangle $, to the first qubit. The resulting state becomes,
\begin{equation}
\left( {{g_0}\left| 0 \right\rangle  + {g_1}\left| 1 \right\rangle } \right)\left| 0 \right\rangle.
\end{equation}

\item Perform the following controlled operation:
\begin{equation}
\left| x \right\rangle \left| 0 \right\rangle  \to \left| x \right\rangle \left( {{\textstyle{{{f_{x0}}} \over {{g_x}}}}\left| 0 \right\rangle  + {\textstyle{{{f_{x1}}} \over {{g_x}}}}\left| 1 \right\rangle } \right) \quad,
\end{equation}
where $x=\{0,1\}$.
\end{enumerate}
The final state is exactly the same as that in Eq.~(\ref{state:2qubit_2}) or Eq.~(\ref{state:2qubit}).

Consider, more generally, the preparation of the following $n$-qubit quantum state~\cite{Zalka1998b,Grover2002,Kaye2004a,Ward2009a}:
\begin{equation}
\sum\limits_{x = 0}^{{2^n} - 1} {f\left( x \right)} \left| x \right\rangle \quad.
\end{equation}
Here again we will assume that $f(x)$ is real. We can image that this is the wavefunction of a particle in 1D. The first qubit describes whether the particle is located in the left half $\left| 0 \right\rangle$ or right half $\left| 1 \right\rangle$ of the line divided by $L \equiv 2^n$ divisions. The first step is therefore to rotate the first qubit as $\cos {\theta _0}\left| 0 \right\rangle  + \sin {\theta _1}\left| 1 \right\rangle$, where
\begin{equation}
\cos^2 {\theta _0} = {{\sum\limits_{0 \le x < L/2} {f{{\left( x \right)}^2}} }} \quad
\end{equation}
represents the probability of locating the particle at the left side, i.e. $0 \le x < L/2$. The next step is to apply the following controlled rotation:
\begin{equation}\label{state:c_rot}
\left| x \right\rangle \left| 0 \right\rangle  \to \left| x \right\rangle \left( {{\textstyle{{\cos {\theta _{x0}}} \over {\cos {\theta _x}}}}\left| 0 \right\rangle  + {\textstyle{{\cos {\theta _{x1}}} \over {\cos {\theta _x}}}}\left| 1 \right\rangle } \right) \quad,
\end{equation}
where 
\begin{equation}
{\cos ^2}{\theta _{00}} = \sum\limits_{0 \le x < L/4} {f{{\left( x \right)}^2}} \quad {\rm{and}}\quad {\cos ^2}{\theta _{01}} = \sum\limits_{L/4 \le x < L/2} {f{{\left( x \right)}^2}} 
\end{equation}
represents the probability for finding the particle in the `$00$' division (${0 \le 0 < L/4}$) and the `$01$' division (${L/4 \le 0 < L/2}$) respectively; an analogous arguments apply for the `$10$' and `$11$' divisions. 
\begin{figure}[t]
\centering
\includegraphics[width=0.55\textwidth]{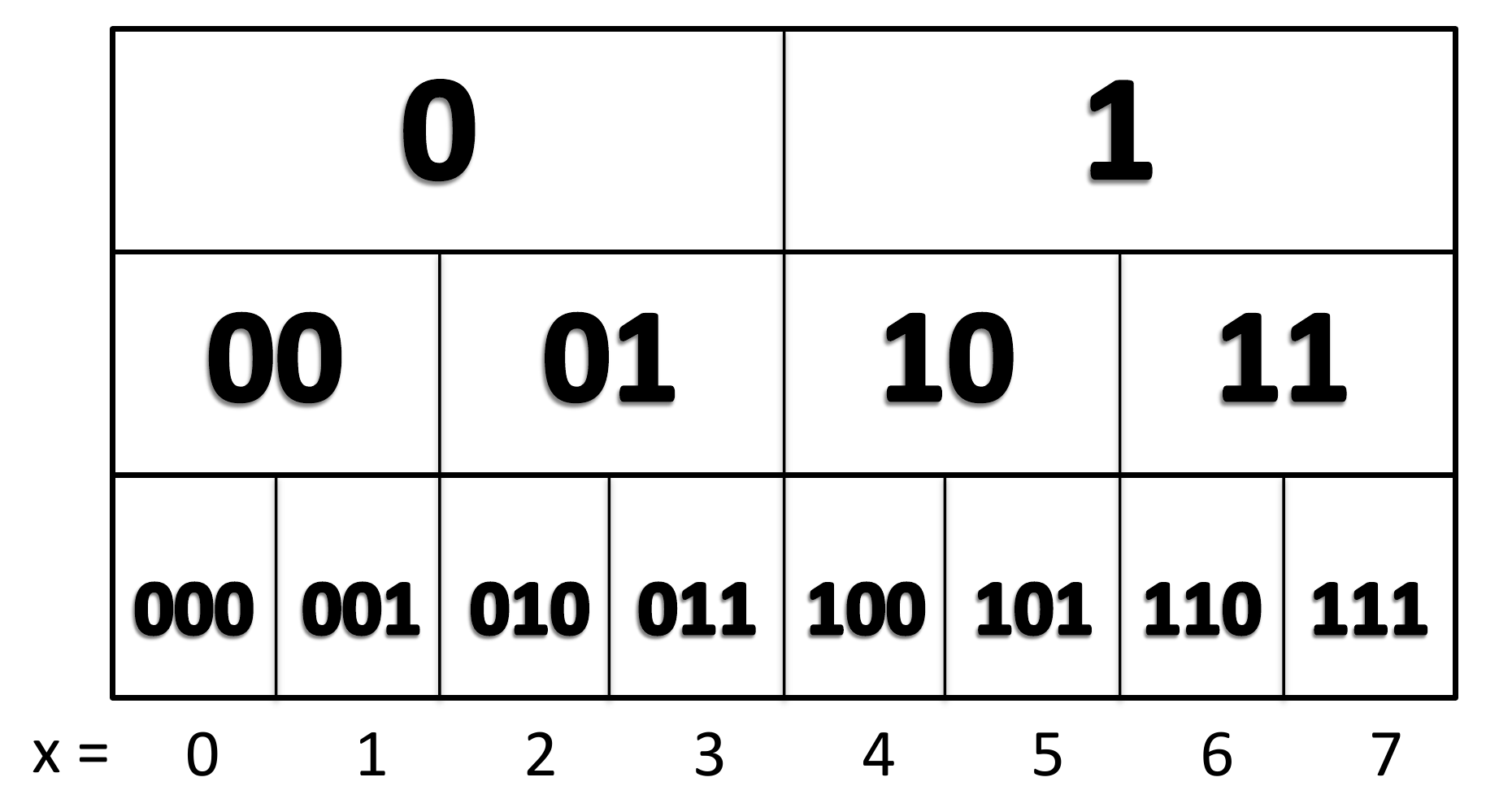}
\caption{Example for the state preparation method. The space is divided in $L=8$ divisions. the `$0$' division refers to the left half of the space, $(0 \le x < L/2)$, and similarly for the `$1$' division. Finer resolution is achieved by increasing the number of labeling digits.}
\label{fig:line}
\end{figure}
In the remaining steps, similar controlled operations described in
Eq.~(\ref{state:c_rot}) are applied, which depend on the division of the controlling qubits. The $\theta$ rotation angles have to be calculated explicitly. It is therefore necessary that the function $f(x)$ is efficiently integrable \cite{Grover2002}. This is expected, as otherwise such a simple algorithm would be able to solve the random-field Ising spin models and other $\mathsf{NP}$-complete problems. We will cover the creation of thermal states later. In the next section, we will consider methods for preparing ground states.

\subsubsection{Preparing ground states}\label{ssec:gr_states}

\subsubsection*{Phase-estimation based methods}
Finding ground states of classical Hamiltonians, e.g. a random-field
Ising model, is known to be ${\mathsf {NP}}$-hard. Therefore, it is
not expected that a quantum computer would be able to solve it
efficiently in general. Furthermore, preparing the ground-state of a
general quantum Hamiltonian $H$ is even more challenging as both
eigenvalues and eigenvectors are required to be obtained, and this
problem belongs to the ${\mathsf {QMA}}$ complexity class, the quantum
analog of ${\mathsf {NP}}$. Fortunately, many problems in physics and
chemistry exhibit structures and symmetries that allow us to arrive at
solutions that are approximation of the exact solutions; for example,
the BCS wavefunction related to superconductivity and superfluidity,
and the Laughlin wavefunction related to the fractional quantum Hall
effect (FQHE), both provide good predictions for the
corresponding many-body problems. The quality of other approximated
solutions, such as the mean-field  or Hartree-Fock approximation,
may vary from problem to problem.


The quality of the approximated solution (or trial solution) ${\left| {{\psi _T}} \right\rangle }$ can be quantified by the fidelity $F$ defined by 
\begin{equation}
F \equiv {\left| {\left\langle {{e_0}} \right.\left| {{\psi _T}} \right\rangle } \right|^2} \quad,
\end{equation}
where $\left| {{e_0}} \right\rangle$ is the target ground state
(assumed unique) of the Hamiltonian $H$ of interest.  The physical
meaning of $F$ is that if one can implement a projective measurement
$\left\{ {\left| {{e_k}} \right\rangle \left\langle {{e_k}} \right|}
\right\}$ in the eigenvector basis $\left\{ {\left| {{e_k}}
    \right\rangle } \right\}$ of $H$ to the trial state ${\left|
    {{\psi _T}} \right\rangle }$, then the probability of getting the
ground state ${\left| {{e_0}} \right\rangle }$ is exactly equal to
$F$, and can be implemented with the phase estimation
algorithm~\cite{Abrams1999}. A similar procedure can produce low
energy eigenstates even if there is no gap~\cite{Poulin2009c}.

With the methods of the previous paragraph, if the projection on the
ground state fails, the initial approximation must be reconstructed
again. Because the projection fails with probability $1-F$, the
approximate preparation must be done $1/(1-F)$ times in average. This
can be improved using phase amplification (a trick similar to Grover's
search) to $\sqrt{1/(1-F)}$ ``coherent'' initial state preparations. A
different method is possible if, as is often the case, we can evolve
with a Hamiltonian $\tilde H$ for which the state approximation is a
ground state. Assume that the approximated ground state has an energy
gap bounded by $\Delta$ for $\tilde H$ and the exact ground state has
a similar gap for $H$. Then we can transform a single preparation of
the approximated state into the exact ground state using around
$1/(1-F)$ phase estimations, each implemented with a time evolution
for a time of $1/\Delta$~\cite{boixo_fast_2010}.

Therefore, a quantum computer, even if it can not solve all
ground-state problems efficiently, is capable to leverage classical
trial states, and solve a border class of problems than those
efficiently solvable by classical computers.

\subsubsection*{Adiabatic state preparation}\label{sec:adiabatic_state_preparation}
The adiabatic method is an alternative way to prepare ground
states~\cite{Farhi2000,Aharonov2003,Perdomo2008b,boixo_eigenpath_2009,Biamonte2010,boixo_fast_2010}. The
original idea is due to \citeasnoun{Farhi2000}. We first must be able to efficiently
prepare the ground state $\left| {\psi \left( 0 \right)}
\right\rangle$ of a certain initial Hamiltonian $H(0) = H_i$. Then we change
the Hamiltonian $H(t)$ slowly, e.g.,
\begin{equation}
H\left( t \right) = \left( {1 - t/T} \right){H_i} + \left( {t/T} \right){H_f} \;.
\end{equation}
Notice that for many reasonable choices of $H_i$ and most physical
Hamiltonians $H_f$ the Hamiltonian $H(t)$ can be simulated using the
methods of Sec.~\ref{sec:simulation_of_time_evolution}. Nevertheless,
common two-body Hamiltonians could be simulated
directly. 

If the change from $H_i$ (when $t=0$) to the target Hamiltonian
$H_f$ (when $t=T$) is slow enough, then the state $\left| {\psi
    \left( t \right)} \right\rangle$, satisfying the time-dependent
Schr\"{o}dinger equation
\begin{equation}
i\hbar {d \over {dt}}\left| {\psi \left( t \right)} \right\rangle  = H\left( t \right)\left| {\psi \left( t \right)} \right\rangle \quad,
\end{equation}
follows the corresponding eigenstate of $H(t)$ adiabatically. This means that $\left| {\psi \left(
      T \right)} \right\rangle$ is close to the ground state of the target Hamiltonian $H_f$. A sufficient
condition for the total time $T$ to ensure the adiabaticicty for a linear interpolation between two Hamiltonians is
\begin{equation}\label{adiabatic:T}
T \gg \frac {\|\partial _sH( s ) \|}  {\Delta _{\min }^2} \quad,
\end{equation}
where $s \equiv t/ T$. Here 
\begin{equation}
{\Delta _{\min }} \equiv \mathop {\min }\limits_{0 \le s \le 1} \left( {{E_1}\left( s \right) - {E_0}\left( s \right)} \right)
\end{equation}
is the minimum gap between the instantaneous eigen-energies $E_1(s)$
and $E_0(s)$ of the first excited state and the ground state. The
following bound has a better dependence on the minimum gap and it also
holds for general (non-linear) interpolations if the rate of change of
the instantaneous eigenstate $| \partial_s \psi(s) \rangle$ is known~\cite{boixo_eigenpath_2009}
\begin{equation}
  T > \frac {\mathcal L^2}{\Delta_{\rm min}} \;.
\end{equation}
Here ${\mathcal L}$ is the path length given by the equation\footnote{More precisely, for this equation we must make a choice of phases such that $\langle \partial_s g(s) | g(s) \rangle$.}
\begin{equation}
  {\mathcal L} = \int \|  \partial_s \psi(s)\rangle \| ds\;.
\end{equation}

Using the methods of Sec.~\ref{sec:simulation_of_time_evolution} adiabatic evolutions can be
simulated efficiently on a quantum circuit. That is, for
cases where one may not be able to physically implement $H(t)$, it is
still possible to turn the adiabatic state preparation into a quantum algorithm and simulate the
adiabatic process in a digital quantum computer. Furthermore, in this
case the total time of the adiabatic evolution can be improved
to\footnote{\citeasnoun{boixo_necessary_2010} have shown that this
  expression for the total evolution time is also optimal.}~\cite{boixo_fast_2010}:
\begin{equation}
  T > \frac {\mathcal L}{\Delta_{\rm min}} \;.
\end{equation}

The remaining question is, in terms of finding ground states, `how
good are adiabatic algorithms?'. As we have seen, the performance, or computational
complexity, of adiabatic algorithms generically depends on the scaling
of the minimal gap $\Delta_{\rm min}$.  Even for classical target
Hamiltonians $H_f$, whether adiabatic algorithms success in solving
$\mathsf{NP}$-problems is still a controversial
issue~\cite{Altshuler2010Anderson-locali,Knysh2010}. Numerical results
suggest that for the classical satisfiability (SAT) problems, the
scaling of the gap would be
exponential~\cite{Young2008,Young2010a,Hen2011}. If the target
Hamiltonian is quantum, the problem is QMA-complete. Nevertheless, we can in principle
apply the adiabatic algorithm to the trial states to improve the
ground-state fidelity~\cite{Oh2008a}, which gives us higher
probability to project into the exact ground state by the phase
estimation algorithm discussed in the previous section.

\subsubsection{Preparing thermal states using quantum Metropolis}
We now consider the preparation of density matrices for thermal states 
\begin{equation}\label{thermal:rho}
\rho _{th}  = e^{ - \beta H} /{\rm Tr}\left( e^{ - \beta H}  \right)\;,
\end{equation}
where $H$ can be a quantum or classical Hamiltonian, and $\beta = 1/T$ is the inverse temperature. We simplify the notation by choosing our units so that the Boltzmann constant $k_B$ is $1$. \citeasnoun{Zalka1998b} and \citeasnoun{Terhal2000b} proposed to simulate the Markovian dynamics of the system by modeling the interaction with a heat-bath by some ancilla qubits. A similar idea has been recently investigated by \citeasnoun{Wang2011}. \citeasnoun{Terhal2000b} also attempted to prepare thermal states by generalizing classical Metropolis-type sampling~\cite{Gould2007}. This first quantum Metropolis algorithm was limited by the fact that it was not possible to control the update rule for the Metropolis algorithm, which would generally lead to a slow convergence rate of the underlying Markov chain. A significant improvement upon this work has been presented recently in \citeasnoun{Temme2009} with the ``quantum Metropolis sampling" algorithm. This algorithm also constructs a Markov-chain whose fixed point is a thermal state Eq. (\ref{thermal:rho}), but the transitions between states can be engineer to achieve faster convergence. The underlying time cost of this algorithm scales as $O(1/{\Delta})$~\cite{Aldous1982}, where $\Delta$ is the eigenvalue gap of the Markov matrix associated with the Metropolis algorithm. 

\citeasnoun{Szegedy2004} introduced a quantum algorithm to speedup classical Markov chains. \citeasnoun{Richter2007} extended this methods to some quantum walks with decoherence. Szegedy's method has also been applied to achieve a quadratic speedup of classical simulated annealing algorithms~\cite{Somma2008,Wocjan2008a}. Further, \citeasnoun{Yung2010c}  achieved a similar speedup of the quantum Metropolis sampling algorithm for quantum Hamiltonians. This algorithm outputs a coherent encoding of the thermal state (CETS):
\begin{equation}
\left| {\psi _{th} } \right\rangle  = \sum\limits_k {\sqrt {e^{ - \beta E_k } /{\mathcal Z}} } \left| {e_k } \right\rangle \quad,
\end{equation}
where $\mathcal Z$ is the partition function, and $E_k$ is the eigenvalue associated with the eigenvector $\left| {e_k } \right\rangle$ of the Hamiltonian $H$. 

Markov chain approaches are practical for many applications. However, for systems like spin glasses, the eigenvalue gap $\Delta$ of the corresponding Markov matrices typically become exponential small, making it inefficient. Several alternative approaches have been already introduced in the literature. Among them: exploiting the transfer-matrix structure of spin systems~\cite{Lidar1997,Yung2010}; mapping the CETS as the ground state of certain Hamiltonian~\cite{Somma2007,Ohzeki2010}; and methods based on quantum phase estimation~\cite{Poulin2009a,Bilgin2010,Riera2011}. In the next section we modified one of these algorithms to prepare thermal states building up from small to bigger subsystems.


\subsubsection{Preparing thermal states with perturbative updates}\label{sec:perturbative_update}
The quantum Metropolis algorithms of the previous subsection extend the advantages of Markov chain Monte Carlo methods to the quantum case, even if we do not know how to diagonalize the quantum Hamiltonian. It is expected that, as in the classical case, they will exhibit good performance for most Hamiltonians. Nevertheless, for very complex systems, such as strongly correlated molecules, it might be difficult to design rules to choose appropriate Markov chain update candidate states, or the convergence rate to the thermal state might still be too slow. In this subsection we will show how, even in the worst case, quantum algorithms for preparing thermal states will exhibit a substantial speedup over classical algorithms, elaborating upon the method of \citeasnoun{Bilgin2010}.

\begin{figure}[t] \centering \includegraphics[width=0.55\textwidth]{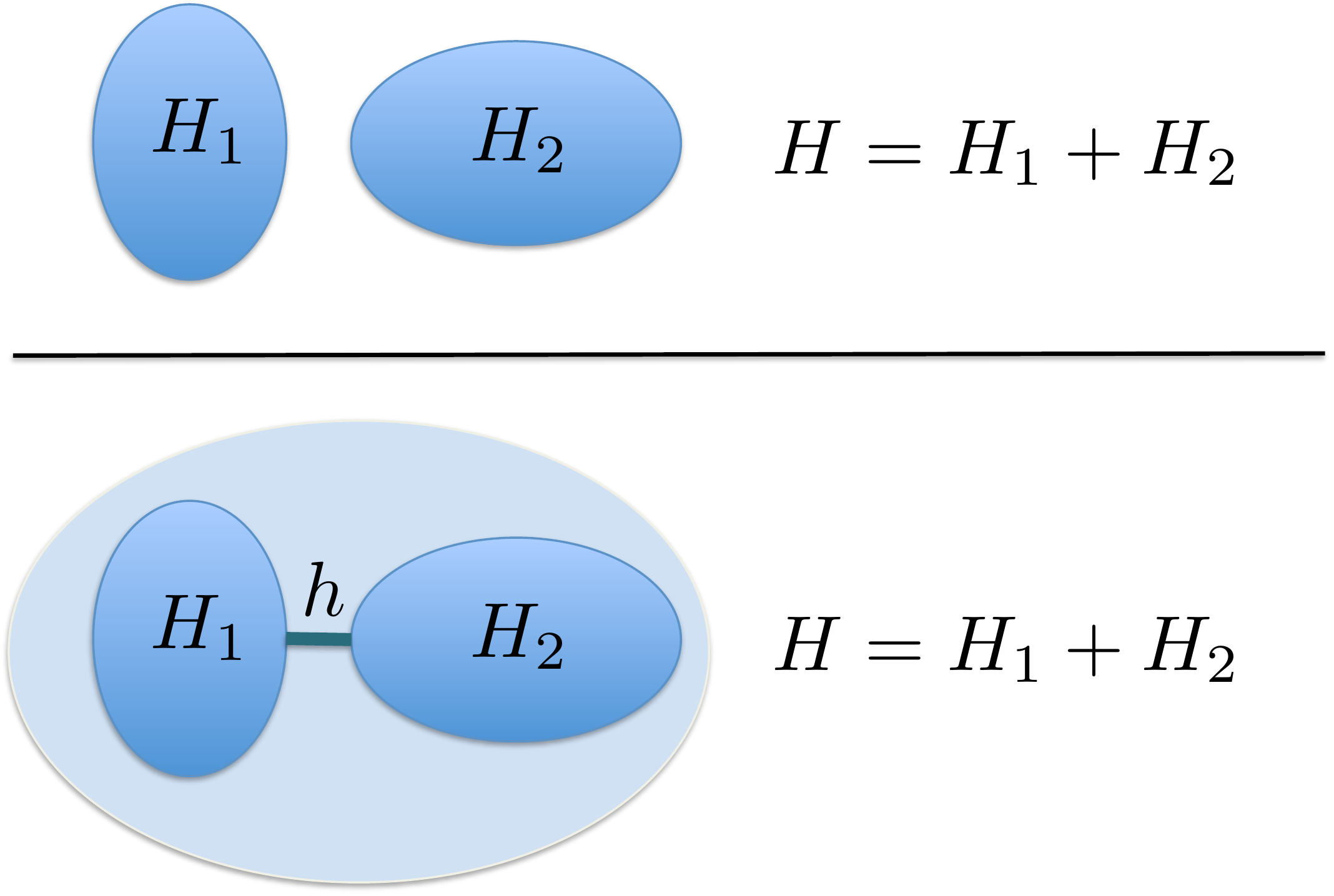} \caption{Pictorial representation of the perturbative update method. The top figure depicts two quantum systems with Hamiltonians $H_1$ and $H_2$ whose thermal states we can prepare (maybe through prior perturbative updates). The bottom figure depicts a single system where the two quantum systems of the top figure have been linked with Hamiltonian $h$. The perturbative update technique is a method to prepare the thermal state of the linked system from thermal states of the smaller subsystems.}  \label{fig:perturbative_update} \end{figure}

The core of this algorithm is a perturbative update subroutine that builds the thermal state $\rho^{(\epsilon)} \propto e^{-\beta (H+\epsilon h)}$ from the state $\rho^{(0)}\propto e^{- \beta H}$. We can use this subroute to build thermal states of complex systems out of thermal states of their components (see Fig.~\ref{fig:perturbative_update}). For this we start with the thermal states of subsystems with Hamiltonians $H_1$ and $H_2$, and use them to prepare the thermal state of the linked system with Hamiltonian $H_1 + H_2 + h$. The coupling Hamiltonian $h$ is introduced perturbatively with a sequence of updates:
\begin{equation}\label{eq:sequence}
  \rho^{(0)} \to \rho^{(\epsilon)} \to \rho^{(2\epsilon)} \to
  \cdots \to \rho^{(1)}\;.
\end{equation}

Quite possibly the thermal states of the smaller subsystems have themselves been prepared with a perturbative update over still smaller pieces.  As in the quantum Metropolis case, it is not necessary to know how to diagonalize the corresponding Hamiltonian.

The perturbative update subroutine is probabilistic, and succeeds with probability $1-\epsilon \beta {\mathrm{Tr\,}} \rho^{(0)} h $, which gives the dominant cost of the algorithm. If the perturbative update fails, we must reconstruct the state $\rho^{(0)} $.  The probability of failure is given by  the maximum change of a thermal state $\rho^{(0)} \propto e^{- \beta H}$ introduced by a perturbation $\epsilon h$, which we now bound. We denote with $Z = {\mathrm{Tr\,}} \rho^{(0)}$ the partition function of $\rho^{(0)}$. Using the Dyson series expansion in imaginary time we write \begin{equation}
 \frac {e^{- \beta (H+\epsilon h) \lambda}} Z = \frac {e^{-\beta H \lambda}} Z -\frac {\epsilon \beta} Z\int_0^\lambda d\lambda_1 \,e^{-\beta H (\lambda-\lambda_1)} h
  e^{- \beta H \lambda_1}  + \ldots \label{eq:dyson_imaginary_time}\;.
\end{equation}
The appropriate measure of the difference between two density matrices is the trace norm $\| \cdot \| _ {\mathrm{Tr\,}} $. The reason is that this norm bounds the difference of arbitrary measurement results for those two states. The trace norm for an arbitrary operator $A$ is given by the sum of the eigenvalues of $\sqrt{A^\dagger A}$, and often scales with the dimension of the operator. We want to do better for the trace norm of the difference between a thermal state and its perturbation, because their dimension grows exponentially (in the number of subsystems). We give such a bound next. 

We will use the following inequality which applies to all unitarily invariant matrix norms~\footnote{See, for instance, Theorem 5.4.7 in \cite{bhatia_positive_2007}.}
\begin{equation}
\Big|\!\Big\| \int_0^1 A^t X B^{1-t} dt \Big\|\!\Big| \le 1/2 |\!\|AX + XB\|\!| \label{eq:bhatia}
\end{equation}
Applying this inequality to the trace norm of the Dyson series of a perturbed thermal state we obtain the bound 
\begin{equation}
(1/Z) \Big\|  \epsilon \beta \int_0^1 d\lambda_1 \,e^{-\beta H (1-\lambda_1)} h
  e^{- \beta H \lambda_1} \Big\| _ {\mathrm{Tr\,}}  \le \epsilon \beta\|h\| \label{eq:firstorderbound}
\end{equation}
where $\|h\|$ is the operator norm of $h$. Notice that the operator norm $\|h\|$ is the highest eigenvalue of $h$, and does not scale with the dimension of $H$ (or even $h$)\footnote{Although strictly speaking we have derived the bound on the change of the thermal state here only to second order, it can be shown to be valid to all orders. For that, we use the exact formula for the perturbation $\epsilon \beta \int_0^1 d\lambda_1 \,e^{-\beta H (1-\lambda_1)} h
  e^{- \beta (H + h)  \lambda_1}$, and the same matrix norm inequality. In addition, we need to use the bound for the change on the partition from ~\citeasnoun{Poulin2009a}. }.

The perturbative update subroutine is composed of two operations. The first operation implements the quantum map
  \begin{equation}
    (\rho^{(0)} ) \to  ( 1- \epsilon \beta h/2) \rho^{(0)} ( 1 - \epsilon \beta h/2)\, {\mathcal{N}}\;,
  \end{equation}
  where ${\mathcal{N}}$ is just a normalization factor.  Similar to the algorithms of the previous section, this map is implemented with phase estimation and a conditional rotation on an ancillary system. The ancillary system is then measured. This measurement can fail, which corresponds to implementing the wrong transformation in the thermal state. The success rate is $1-\epsilon \beta {\mathrm{Tr\,}} \rho^{(0)} h $. When the measurement of the ancilla system fails, the thermal state can not be recovered, and we must start again from the beginning. The cost of the phase estimation is $\epsilon^{-1} \beta^{-1} \|h\|^{-2}$. This operation can be understood as an update of the Gibbs probabilities of $\rho^{(0)} $ to those of $\rho^{(\epsilon)} $. The second operation of the perturbative update is a transformation to the eigenbasis of $\rho^{(\epsilon)} $. This is implemented by ``dephasing'' in that basis, which is achieved by evolving for a random amount of time (with expectation time $\epsilon \|h\|$) using the Hamiltonian $H + \epsilon h$. This completes the perturbative update subroutine.



\subsection{Algorithmic quantum cooling}\label{Qcooling}
\citeasnoun{Yung2011AQC} presented an algorithmic quantum cooling approach that transforms any input state $\rho_{in}$ into an output state $\rho_{out}$ which is guaranteed to have lower energy with respect to a given Hamiltonian $H$. Explicitly,
\begin{equation}
{\rm Tr}\left( {H\rho _{out} } \right) < {\rm Tr}\left( {H\rho _{in} } \right) \quad .
\end{equation}
In principle, this algorithm can cool the resulting quantum state to a state arbitrarily close to the ground state of $H$. Nevertheless, like the ground-state algorithms of Sec.~\ref{ssec:gr_states}, the efficiency is related to the energy gap $\Delta$ between the ground state and the excited state(s). Depending on how the algorithm is implemented, this dependence can scale like $O(1/\Delta^2)$ or $O(1/\Delta)$.

Algorithmic quantum cooling first entangles an ancilla qubit with the system state. When the ancilla qubit is measured, a result of $\left| 0 \right\rangle$ correlates with a cooler system state. On average, however, there is no gain or loss of energy. This measurement is used to gain information, just like a Maxwell's demon. 
The measurement outcome of the ancilla qubit in algorithmic quantum cooling can be mapped into a 1D random walk. The walker starts at $x=0$. For the cooling outcome, the walker makes a step towards the positive side $x>0$, and towards the negative side $x<0$ for the heating outcome. If the walker moves too far to the negative side, the procedure is restarted. For some range of parameters, whenever the walker goes to the negative side $x<0$, it is guaranteed that the quantum state is hotter than the original state. Therefore, removing these hot walkers will reduce the average energy over an ensamble of walkers, just like in evaporative (or ``coffee") cooling of gas molecules.  The procedure stops once the walker has moved sufficiently to the positive side.

\subsubsection{Basic idea of the quantum cooling method}
\begin{figure}[t]
\centering
\includegraphics[width=0.7\textwidth]{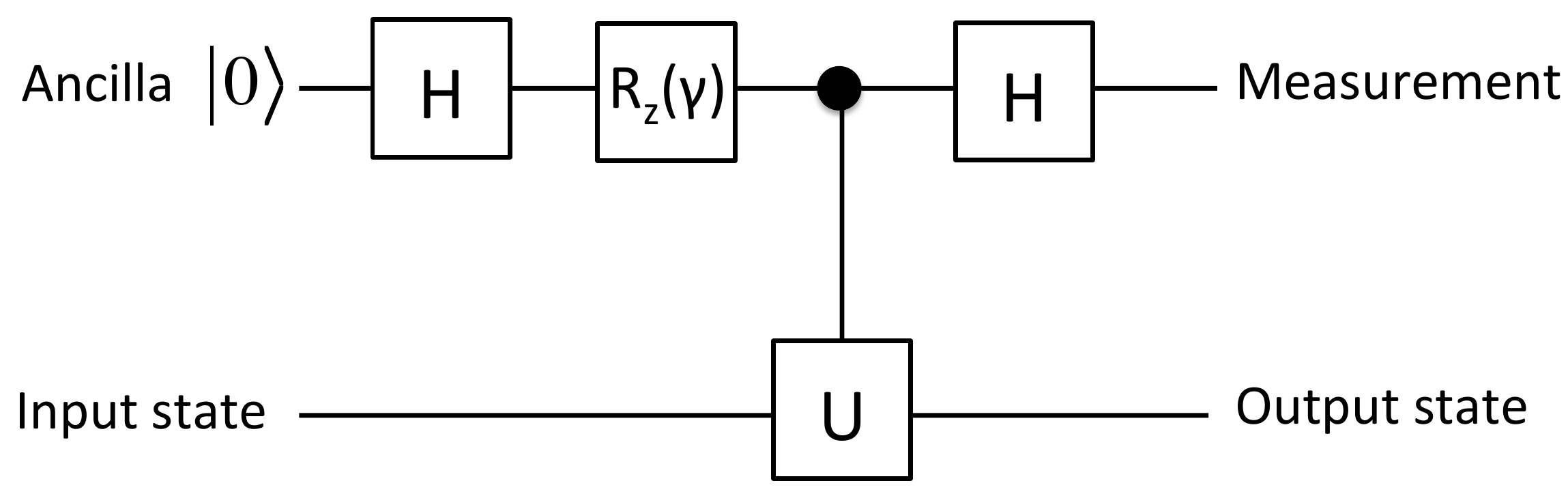}
\caption{Quantum circuit diagram for the quantum cooling algorithm. Here $\mathsf{H} = {\textstyle{1 \over {\sqrt 2 }}}\left( {\left| 0 \right\rangle  + \left| 1 \right\rangle } \right)\left\langle 0 \right| + {\textstyle{1 \over {\sqrt 2 }}}\left( {\left| 0 \right\rangle  - \left| 1 \right\rangle } \right)\left\langle 1 \right|$ is the Hadamard gate, $\mathsf{R_z} \left( \gamma  \right) = \left| 0 \right\rangle \left\langle 0 \right| - ie^{i\gamma } \left| 1 \right\rangle \left\langle 1 \right|$ is a local phase gate, and $U\left( t \right) = e^{ - i H_s t}$, is the time $t$ evolution operator simulating the dynamics of the system under the Hamiltonian $H_s$.}
\label{fig:cooling_circuit}
\end{figure}
We now sketch the basic working mechanism of algorithmic quantum cooling. The core component of this cooling algorithm consists of four quantum gates (see Fig.~\ref{fig:cooling_circuit})\footnote{Similar quantum circuits are used in DQC1 and phase estimation, for instance.}. The first gate is a so-called Hadamard gate:
\begin{equation}
\mathsf{H} \equiv {\textstyle{1 \over {\sqrt 2 }}}\left( {\left| 0 \right\rangle  + \left| 1 \right\rangle } \right)\left\langle 0 \right| + {\textstyle{1 \over {\sqrt 2 }}}\left( {\left| 0 \right\rangle  - \left| 1 \right\rangle } \right)\left\langle 1 \right| \quad.
\end{equation}
It is followed by a local phase gate, 
\begin{equation}
\mathsf{R_z} \left( \gamma  \right) \equiv \left| 0 \right\rangle \left\langle 0 \right| - ie^{i\gamma } \left| 1 \right\rangle \left\langle 1 \right| \quad,
\end{equation}
where the parameter $\gamma$ plays a role in determining the overall efficiency of the cooling performance of the algorithm. The interaction with the Hamiltonian $H$, which can be either quantum or classical, is encoded in the time evolution operator 
\begin{equation}
U\left( t \right) = e^{ - i H_s t}\;.
\end{equation}
As explained above, time evolution can be implemented efficiently in a quantum computer. 

The operation of the circuit in Fig.~\ref{fig:cooling_circuit} on input state $\left| {\psi _{in} } \right\rangle $ is as follows. 
\begin{description}
\item[Step 1] State initialization,
\begin{equation}
\left| {\psi _{in} } \right\rangle \left| 0 \right\rangle \quad,
\end{equation}
with the ancilla state in $\left| 0 \right\rangle$.

\item[Step 2] Apply the Hadamard gate, $\mathsf{H} = {\textstyle{1 \over {\sqrt 2 }}}\left( {\left| 0 \right\rangle  + \left| 1 \right\rangle } \right)\left\langle 0 \right| + {\textstyle{1 \over {\sqrt 2 }}}\left( {\left| 0 \right\rangle  - \left| 1 \right\rangle } \right)\left\langle 1 \right|$, and the local phase gate $\mathsf{R_z} \left( \gamma  \right) = \left| 0 \right\rangle \left\langle 0 \right| - ie^{i\gamma } \left| 1 \right\rangle \left\langle 1 \right|$ to the ancilla qubit,
\begin{equation}
\left| {\psi _{in} } \right\rangle \left( {\left| 0 \right\rangle  - i e^{i \gamma} \left| 1 \right\rangle } \right)/\sqrt 2 \quad .
\end{equation}

\item[Step 3] Apply the controlled-$U(t)$ to the system state,
\begin{equation}
\left( {\left| {\psi _{in} } \right\rangle \left| 0 \right\rangle  - ie^{i\gamma } U\left( t \right)\left| {\psi _{in} } \right\rangle \left| 1 \right\rangle } \right)/\sqrt 2 \quad .
\end{equation}

\item[Step 4] Apply the Hadamard to the ancilla qubit again. This produces the following output state:
\begin{equation}\label{core_step}
\Lambda _0 \left| {\psi _{in} } \right\rangle \left| 0 \right\rangle  + \Lambda _1 \left| {\psi _{in} } \right\rangle \left| 1 \right\rangle \quad,
\end{equation}
\end{description}
where $\Lambda _j  \equiv \left( {I + (-1)^{j+1} ie^{i\gamma } U} \right)/2$ for $j=\{0,1\}$. 

A projective measurement on the ancilla qubit in the computational basis $\left\{ {\left| 0 \right\rangle ,\left| 1 \right\rangle } \right\}$ yields one of the two (unnormalized)  states, 
\begin{equation}
\left( {I \pm ie^{i\gamma } U} \right)\left| {\psi _{in} } \right\rangle
\end{equation}
Their mean energy is either higher (for outcome $\left| 1 \right\rangle$, $x$ is decreased by 1) or lower (for outcome $\left| 0 \right\rangle$, $x$ is increased by 1) than that of the initial state $\left| {\psi _{in} } \right\rangle $. 

To justify this assertion, let us expand the input state, 
\begin{equation}
\left| {\psi _{in} } \right\rangle  = \sum\nolimits_k {c_k } \left| {e_k } \right\rangle \quad,
\end{equation}
in the eigenvector basis $\left\{ {\left| {e_k } \right\rangle } \right\}$ of the Hamiltonian $H$. Note that 
\begin{equation}
\left| {\left( {1 \pm ie^{i\gamma } U} \right)\left| {e_k } \right\rangle } \right|^2  = \left| {c_k } \right|^2 \left( {1 \pm \sin \phi _k } \right) \quad,
\end{equation}
where $\phi _k  \equiv E_k t - \gamma$ depends on the eigen-energy $E_k$ of $H$. For simplicity, we will assume that one can always adjust the two parameters, $\gamma$ and $t$, such that 
\begin{equation}
 - {\textstyle{\pi  \over 2}} \le \phi _k  < {\textstyle{\pi  \over 2}}
\end{equation}
for all non-negative integers $k$. Then, the factors $(1 - \sin \phi _k)$ are in descending order of the eigen-energies, and the opposite is true for the factors $(1 + \sin \phi _k)$. Therefore, apart from an overall normalization constant, the action of the operator $({I \pm ie^{i\gamma } U})$ is to scale each of the probability weights $\left| {c_k } \right|^2$ by an eigen-energy dependent factor $(1 \pm \sin \phi _k)$, i.e.,
\begin{equation}\label{scaling_factor}
\left| {c_k } \right|^2  \to \left| {c_k } \right|^2 \left( {1 \pm \sin \phi _k } \right) \quad.
\end{equation}
The probability weights  scale to larger values,  i.e., 
\begin{equation}
\left( {1 - \sin \phi _k } \right)/\left( {1 - \sin \phi _j } \right) > 1
\end{equation}
for the eigen-energy $E_k < E_j$ in the cooling case (i.e., for outcome $\left| 0 \right\rangle$), and vice versa for the heating case (i.e., for outcome $\left| 1 \right\rangle$). Further cooling can be achieved by applying the quantum circuit repeatedly and reject/recycle the random walker when $x<0$.

\subsubsection{Connection with heat-bath algorithmic cooling}
The  algorithmic quantum cooling approach is related to the well-known heat-bath algorithmic cooling (HBAC) \cite{Boykin2002,Baugh2005,Schulman2005}. HBAC aims to polarize groups of spins as much as possible, i.e. to prepare the state 
\begin{equation}
\left| { \uparrow  \uparrow  \uparrow ... \uparrow } \right\rangle \quad.
\end{equation}
This state is important for providing fresh ancilla qubits for quantum error correction as well as for NMR quantum computation. 
In HBAC, some reversible operations are first performed to redistribute the entropy among a group of spins. Some of the spins will become more polarized. For a closed system, there is a so-called Shannon bound~\cite{Schulman2005} which limits the compression of the entropy. In order to decrease the entropy of the whole system, the depolarized spins interact with a physical heat bath that acts as an entropy sink. We note that from an algorithm point of view, the existence of a physical heat bath can be replaced by the (imperfect) preparation of polarized spins by other methods. 
The method of algorithmic quantum cooling from \citeasnoun{Yung2011AQC} may be considered as a generalization of the HBAC, as it is applicable to cool {\it any} physical system that is simulable by a quantum computer, not just non-interacting spins.

\section{Special topics}

\subsection{Adiabatic non-destructive measurements}\label{chp:AQS}

In Section.~\ref{sec:state_preparation} we reviewed several methods to
prepare ground states and thermal states of quantum systems of
interest in physics and chemistry. In particular, in
subsection~\ref{sec:adiabatic_state_preparation} we gave an overview
of the adiabatic method for preparing ground states.  The adiabatic
model may be naturally more robust against noise, offering a method to
perform small to medium size simulations without using sophisticated
error correction schemes. Because of this and other reasons, adiabatic
based quantum computation is possibly easier to realize physically
than quantum computation based on the circuit model. In this section
we review a method to effect non-destructive measurements of constants
of the motion within the adiabatic model.

As explained in subsection~\ref{sec:adiabatic_state_preparation}, it
is in principle possible to adiabatically prepare the ground state of
a physical or chemical system with Hamiltonian $H_f$. There we said
that this can be done by interpolating slowly enough between a simple
initial Hamiltonian $H_i$ and the final Hamiltonian $H_f$. Following
\citeasnoun{Biamonte2010},   we now
add an ancillary qubit subsystem with orthonormal basis
$\{|p_0\rangle, |p_1\rangle\}$. This auxiliary system will be use
for the adiabatic non-destructive measurements. During the adiabatic ground state
preparation, this subsystem is acted upon with Hamiltonian $\delta |p_1
\rangle \langle p_1|$, and therefore it remains in the state %
$|p_0\rangle$. The choice of $\delta > 0$ will be explained shortly. 

The measurement procedure begins by bringing the ancillary qubit and
the system being simulated
into interaction, adiabatically\footnote{The interaction Hamiltonian
  is typically a three-body Hamiltonian, which makes direct
  simulations more difficult. This difficulty can be overcome using
  gadgets~\cite{kempe_complexity_2006,oliveira_complexity_2005,biamonte_realizable_2008,Biamonte2010} or the average Hamiltonian method~\cite{Waugh68}}. We choose the interaction
Hamiltonian $H_{int}=A\otimes|p_1
\rangle \langle p_1|$. Here $A$ is any observable
corresponding to a constant of the motion, that is $[A,H]=0$.  In
particular, the Hamiltonian $H_f$ itself can be used to obtain the
ground state energy.
The total Hamiltonian becomes
\begin{equation}
H_f + \delta |p_1
\rangle \langle p_1| + \underbrace{A \otimes |p_1\rangle\langle p_1|}_{H_{SP}}\;.
\end{equation}
If the energy bias $\delta$ is bigger than the expectation value of
the observable $A$, the state does not change during this initial
interaction~\cite{Biamonte2010}.

After the initial
interaction, we apply a Hadamard gate 
to the ancillary qubit. We denote the time at which we apply this gate
as $t=0$. Let $|s_0\rangle$ be the ground state of $H_f$. After a
further time $t$ the system plus ancilla qubit evolves to 
\begin{equation}
|\psi(t)\rangle = \frac{1}{\sqrt{2}} |s_0\rangle \otimes (|p_0\rangle +
e^{-i  \omega t}|p_1\rangle)
\end{equation}
where $\omega = (a_0+\delta)/\hbar$, and $a_0 = \langle s_0| A |s_0\rangle$ 
is the expectation value we wish to measure. Finally, we again apply a Hadamard gate to the probe.
The resulting state is
\begin{equation}\label{eqn:phase}
|\psi(t)\rangle = |s_0\rangle \otimes \left(\cos \left(\omega t /2\right)
|p_0\rangle+ i\sin\left(\omega t /2\right) |p_1 \rangle \right),
\end{equation}
yielding probability,
\begin{equation}
\label{eqn:prob}
P_0(t) = \frac{1}{2}\left(1 + \cos(\omega t)\right) =
\cos^2\left(\omega t /2\right).
\end{equation}
Measuring the probe does not disturb the state of the simulator which
can be reused for another measurement.  This measurement can be
repeated until sufficient statistics have been accumulated to
reconstruct~$\omega$. We refer to \citeasnoun{Biamonte2010} for
details on numerical simulations and considerations of the influence
of noise.

\subsection{TDDFT and quantum simulation}
Density Functional Theory (DFT) and its time-dependent extension (TDDFT) have become arguably the most widely used methods in computational chemistry and physics. In DFT and TDDFT, the properties of a many-body system can be obtained as functionals of the simple one-electron density rather than the correlated many-electron wavefunction. This represents a great conceptual leap from usual wavefunction-based methods such as Hartree-Fock, configuration interaction and coupled cluster methods and therefore the connections between DFT/TDDFT and quantum computation have just begun to be explored. Since TDDFT is a time-dependent theory, it is more readily applicable to quantum simulation than DFT which is strictly a ground state theory. For recent developments in the connections between DFT and quantum complexity\footnote{It turns out that finding a universal funcional for DFT is QMA hard. \citeasnoun{liu_quantum_2007} proved a related results: $N$-representability is also QMA hard.} see \citeasnoun{Schuch09}, while for applications of DFT to adiabatic quantum computation see \citeasnoun{Gaitan2009}. In this section, we provide a brief overview of the fundamental theorems of TDDFT, which establish its use as a tool for simulating quantum many-electron atomic, molecular and solid-state systems and we mention recent extensions of TDDFT to quantum computation~\cite{tempel_2012}. 

In its usual formulation, TDDFT is applied to a system of N-electrons described by the Hamiltonian
\begin{equation}
\hat{H}(t) = \sum_{i=1}^N \frac{\hat{p}^2_i}{2m} + \sum_{i<j}^N w(|\hat{\mathbf{r}}_i - \hat{\mathbf{r}}_j|) + \int  v(\mathbf{r}, t) \hat{n}(\mathbf{r}) d^3 \mathbf{r},
\label{electron_Hamiltonian}
\end{equation}
where $\hat{\mathbf{p}}_i$ and $\hat{\mathbf{r}}_i$ are respectively the position and momentum  operators of the $i$th electron, $w(|\hat{\mathbf{r}}_i - \hat{\mathbf{r}}_j|)$ is the electron-electron repulsion and $v(\mathbf{r}, t)$ is a time-dependent one-body scalar potential which includes the potential due to nuclear charges as well as any external fields. The electron-electron repulsion, $w(|\hat{\mathbf{r}}_i - \hat{\mathbf{r}}_j|)$, leads to an exponential scaling of the Hilbert space with system-size and makes simulation of the many-electron Schr{\"o}dinger equation on a classical computer intractable.
 $\hat{n}(\mathbf{r}) = \sum_i^N \delta(\mathbf{r} - \hat{\mathbf{r}}_i)$ is the electron density operator, whose expectation value yields the one-electron probability density, $\langle \hat{n}(\mathbf{r}) \rangle \equiv n(\mathbf{r}, t)$. The basic theorems of TDDFT prove that, in principle, one can simulate the evolution of the Hamiltonian in Eq.~\ref{electron_Hamiltonian} using $n(\mathbf{r}, t)$ \textit{directly} and thereby avoid calculating and storing the exponential amount of information in the many-electron wavefunction.

The first basic theorem of TDDFT, known as the ``Runge-Gross" (RG) theorem"~\cite{Runge1984}, establishes the existence of a one-to-one mapping between the expectation value of $\hat{n}(\mathbf{r})$ and the scalar potential $v(\mathbf{r}, t)$. i.e
\begin{equation}
n(\mathbf{r}, t) \leftrightarrow v(\mathbf{r}, t).
\end{equation}
However, $v(\mathbf{r}, t)$ is the only part of the Hamiltonian in Eq.~\ref{electron_Hamiltonian} that is non-universal. i.e. $\sum_{i=1}^N \frac{\hat{p}^2_i}{2m} + \sum_{i<j}^N w(|\hat{\mathbf{r}}_i - \hat{\mathbf{r}}_j|)$ is the same operator for each electronic system. Therefore, due to the uniqueness of the solution to the time-dependent Schr{\"o}dinger equation, the RG theorem establishes a one-to-one mapping between the density and the wavefunction. This implies that the wavefunction is in fact a unique functional of the density,
\begin{equation}
\psi(\mathbf{r}_1,..., \mathbf{r}_N; t) \equiv \psi[n](\mathbf{r}_1,..., \mathbf{r}_N; t),
\end{equation}
as is any observable of the system. The RG theorem implies the remarkable fact that the one-electron density contains the same quantum information as the many-electron wavefunction. This means that in principle, if one had a means of directly simulating $n(\mathbf{r}, t)$, one could extract all observables of the system without ever needing to simulate the many-body wavefunction.

The second basic TDDFT theorem is known as the ``van Leeuwen (VL) theorem"~\cite{Leeuwen99}. It gives an analytic expression for a time-dependent one-body scalar potential that applied to another system with a different, and possibly simpler, electron-electron repulsion $w'(|\hat{\mathbf{r}}_i - \hat{\mathbf{r}}_j|)$, gives the same density evolution as the original Hamiltonian of Eq.~\ref{electron_Hamiltonian}. When $w'(|\hat{\mathbf{r}}_i - \hat{\mathbf{r}}_j|)=0$, this auxiliary system is referred to as the ``Kohn-Sham system"~\cite{Kohn65}. Due to it's simplicity and accuracy, the Kohn-Sham system is in practice used in most DFT and TDDFT calculations. Since the Kohn-Sham system is non-interacting, its wavefunction is simply described by a Slater determinant of single-electron orbitals, which satisfy the time-dependent Kohn-Sham equations,
\begin{equation}
\imath \frac{\partial}{\partial t} \phi_i(\mathbf{r}, t) = \left[ -\frac{1}{2} \nabla^2 + v_{ks}[n](\mathbf{r}, t) \right] \phi_i(\mathbf{r}, t)\;.
\label{kohn_sham}
\end{equation}
The true interacting density is obtained from the orbitals by square-summing; that is, $n(\mathbf{r}, t) = \sum_i |\phi_i(\mathbf{r}, t)|^2$. Naturally, the set of single-particle equations in Eq.~\ref{kohn_sham} are far easier to solve than evolution under the Hamiltonian in Eq.~\ref{electron_Hamiltonian}. In practice, the Kohn-Sham potential, $v_{ks}[n](\mathbf{r}, t)$, must be approximated as a density functional, but the VL theorem rigorously guarantees its existence and uniqueness. 


\begin{figure}[t]
\begin{centering}
\includegraphics[width=1.0\columnwidth]{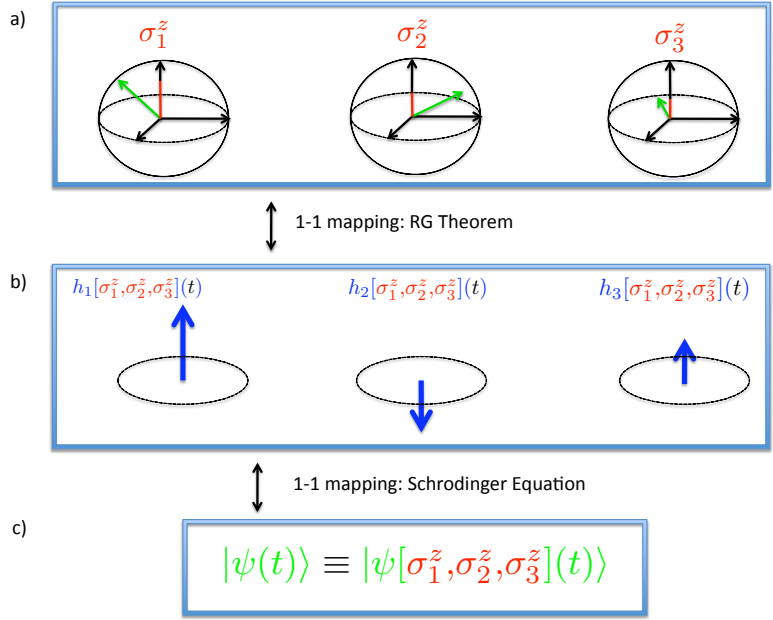} 
\end{centering}
\caption{\textbf{Runge-Gross theorem for a 3 qubit example} - The set of expectation values $\{ \sigma_1^z , \sigma_2^z , ... \sigma_N^z  \}$, defined by the the Bloch vector components of each qubit along the  z-axis in (a), is uniquely mapped onto the set of local fields $\{ h_1, h_2, ... h_N \}$ in (b) through the RG theorem. Then, through the Schr{\"o}dinger equation, the set of fields is uniquely mapped onto the wavefunction. These two mappings together imply that the N-qubit wavefunction in (c) is in fact a unique functional of the set of expectation values $\{ \sigma_1^z , \sigma_2^z , ... \sigma_N^z  \}$.}
\label{RG_theorem}
\end{figure}

\citeasnoun{tempel_2012} recently extended the RG and VL theorems to systems of interacting qubits described by the class of universal 2-local Hamiltonians
\begin{equation}
\hat{H}(t) = \sum_{i=1}^{N-1} J^{\perp}_{i, i+1} (\hat{\sigma}_i^{x} \hat{\sigma}_{i+1}^{x} + \hat{\sigma}_i^{y} \hat{\sigma}_{i+1}^{y}) + \sum_{i =1}^{N-1} J^{\parallel}_{i, i+1} \hat{\sigma}_i^{z} \hat{\sigma}_{i+1}^{z} + \sum_{i=1}^N h_i(t) \hat{\sigma}_i^z\;.
\label{qubit Hamiltonian}
\end{equation}
This Hamiltonians apply to a variety of different systems, particularly in solid-state quantum computing. \citeasnoun{Bose2003} and \citeasnoun{Bose2004} have shown that any set of universal two-qubit and single-qubit gates can be implemented with the Hamiltonian of Eq.~\ref{qubit Hamiltonian}, and therefore it can be used to perform universal quantum computation. The RG theorem applied to such universal Hamiltonians establishes a one-to-one mapping between the set of local fields $\{ h_1, h_2, ... h_N \}$ used to implement a given computation, and the set of single-qubit expectation values $\{ \sigma_1^z , \sigma_2^z , ... \sigma_N^z  \}$ (see Fig.~\ref{RG_theorem}). This implies that one can use single-qubit expectation values as the basic variables in quantum computations rather than wavefunctions and \textit{directly} extract all observables of interest with only knowledge of the spin densities. Naturally, certain properties such as entanglement will be difficult to extract as functionals of the set of spin densities $\{ \sigma_1^z , \sigma_2^z , ... \sigma_N^z  \}$. Nevertheless, \citeasnoun{tempel_2012} give an explicit entanglement functional.
\begin{figure}[t]
\begin{centering}
\includegraphics[width=1.0\columnwidth]{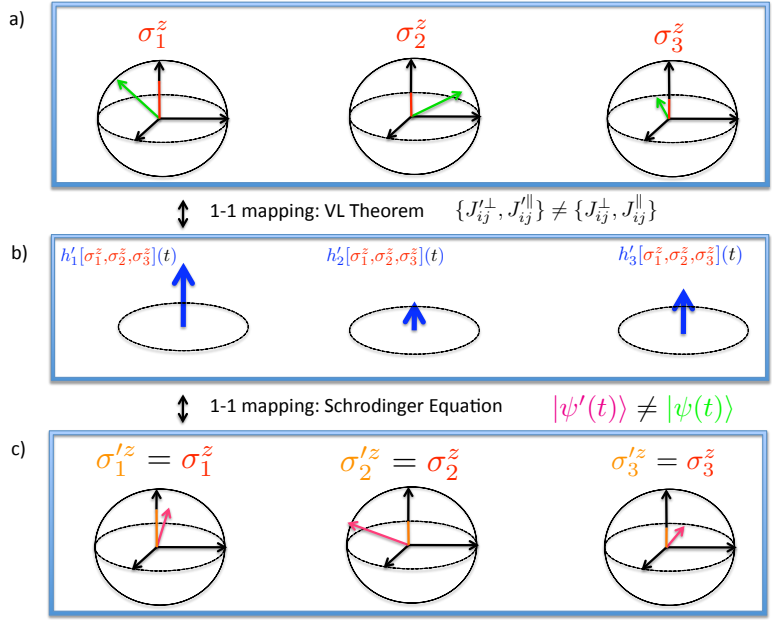} 
\end{centering}
\caption{\textbf{Van Leeuwen theorem for a 3 qubit example} - The set $\{ \sigma_1^z , \sigma_2^z , ... \sigma_N^z  \}$ (a) obtained from evolution under Eq.~\ref{qubit Hamiltonian}, is uniquely mapped to a new set of fields $\{ h'_1, h'_2, ... h'_N \}$ (b) for a Hamiltonian with different two-qubit interactions. Evolution under this new Hamiltonian returns the same expectation values $\{ \sigma_1^z , \sigma_2^z , ... \sigma_N^z  \}$, although the wavefunction is different and hence projections of the Bloch vectors along other axes are in general different (c).}
\label{VL_theorem}
\end{figure}

In addition to the RG theorem, one can derive a VL theorem for qubits. The VL theorem for qubits provides an exact prescription for simulating universal Hamiltonians with other universal Hamiltonians that have different, and possibly easier-to-realize, two-qubit interactions. In analogy to the Kohn-Sham system in electronic TDDFT, one can consider an auxiliary Hamiltonian, 
\begin{eqnarray}
\hat{H}'(t) &=& \sum_{i=1}^{N-1} J'^{\perp}_{i, i+1} (\hat{\sigma}_i^{x} \hat{\sigma}_{i+1}^{x} + \hat{\sigma}_i^{y} \hat{\sigma}_{i+1}^{y}) + \sum_{i =1}^{N-1} J'^{\parallel}_{i, i+1} \hat{\sigma}_i^{z} \hat{\sigma}_{i+1}^{z} \nonumber \\&+& \sum_{i=1}^N h'_i[\sigma_1^z , \sigma_2^z, ... \sigma_N^z](t) \hat{\sigma}_i^z,
\label{qubit Hamiltonian 2}
\end{eqnarray}
with simpler two-qubit couplings $\{J'^{\perp}, J'^{\parallel}\}$. The VL theorem guarantees the existence of the auxiliary fields $\{ h'_1, h'_2, ... h'_N \}$ as functionals of the spin densities which reproduce any given set $\{ \sigma_1^z , \sigma_2^z , ... \sigma_N^z  \}$ that one might wish to simulate. In this way, one can construct an entire class of density functionals for quantum computing that map between different universal Hamiltonians as illustrated in figure~\ref{VL_theorem}.

TDDFT applied to universal qubit Hamiltonians provides a potentially powerful tool for simulating quantum computations on classical computers, similar to how it has been applied in computational chemistry for simulating electronic systems. By choosing the auxiliary "Kohn-Sham" system to be less entangled than the original system, one can hope to simplify simulations of quantum algorithms by finding simple approximations to the auxiliary local fields $\{ h'_1, h'_2, ... h'_N \}$ as functionals of the spin density. As in electronic TDDFT, the development of approximate density functionals for qubit systems will be a necessary next step, which is discussed in \citeasnoun{tempel_2012}.
\section{Conclusion and outlook}


\begin{table}[t!]
\caption{A brief survey of recent experiments on digital quantum simulation}
\scalebox{0.8}{
\begin{tabular}{ p{2.3in}p{3in}p{0.5in} }
\toprule
  {\bf Physical implementations} & {\bf What is simulated?}  & {\bf Scale}\\
  \hline
  & & \\
  Nuclear Magnetic Resonance (NMR)  & $\bullet$ Thermal states of a frustrated magnet ~\cite{Zhang2011} & {4 qubits} \\
  & $\bullet$ Ground state of a pair of interacting Heisenberg spins subject to simulated external fields~\cite{Li2011}& {3 qubits} \\
  & $\bullet$ Isomerization reaction dynamics~\cite{Lu2011} & {3 qubits} \\
  & $\bullet$ Ground state of hydrogen molecule H$_2$~\cite{Du2010}& {2 qubits} \\ 
  & & \\
  Trapped Ions & $\bullet$ Time dynamics of spin systems~\cite{Lanyon2011} & {6 qubits}\\
  & $\bullet$  Dissipative open-system dynamics~\cite{Barreiro2011} & {5 qubits} \\
  & & \\
  Quantum Optics & $\bullet$ 1D quantum walk of a topological system~\cite{Kitagawa2011}  & {4 steps}\\
  & $\bullet$ 1D quantum walk with tunable decoherence~\cite{Broome2010} &  {6 steps} \\
  &  $\bullet$ Ground state of hydrogen molecule H$_2$~\cite{Lanyon2010} & {2 qubits}\\
  & & \\
  \bottomrule
\end{tabular}}
\label{table}
\end{table}%
To the best of our knowledge, the first quantum simulation experiment
was performed by \citeasnoun{Somaroo1999a} in a two-qubit NMR system
in 1999, where a truncated quantum harmonic or anharmonic oscillator
(4 levels) was simulated. Strictly speaking, this belongs to an
``analog" simulation; because the Hamiltonian of the quantum
oscillator was directly simulated by the Hamiltonian of the spins,
instead of applying quantum algorithms. The progress of quantum
simulation is still gaining momentum.  In Table~\ref{table}, we list
several recent experiments on digital quantum simulation. Earlier
references may be found from them and \citeasnoun{Buluta2009a},
\citeasnoun{Kassal2011}, and \citeasnoun{Jones2011}. 

So far, none of the simulations implement active error correction. An important aspect of digital quantum simulation is the resource estimation when fault-tolerant structures are considered. 
We refer to \citeasnoun{Brown2006} and \citeasnoun{Clark2009} for further explore this area. In short, to achieve large-scale quantum simulation, there are still many technological challenges to overcome. For example, in many currently available setups, the performances of the two-qubit gates are still too noisy for fault-tolerant simulation. However, by experimenting with small-scale quantum simulation experiments, we believe that valuable lessons can be learnt to optimize the performance of the currently available technology. 

A related question is ``what is the best way to implement quantum simulation?". For classical computers, there is no doubt that silicon-based semiconductors work successfully. For quantum computers, a general feature of the currently proposed technologies, such as quantum dots, quantum optics, trapped ions, nuclear and electron spins, impurity, superconducting devices etc., is that there is a trade-off between controllability and reliability. Usually, systems that can be controlled easily suffer more from decoherence from the environment. There are two approaches to tackle this problem: one may either look for new systems that are good for both control and can be isolated from the environment, or develop hybrid structures that combine the advantages from both sides. For example, there has been progress in coupling superconducting devices with spin ensembles~\cite{Duty_physics}. The former provides the controllability and the latter provides reliability. In short, the future of quantum computation and quantum simulation is still full of challenges and opportunities. We hope this article can stimulate more ideas that can help move this field forward.

\section{Acknowledgements}
The authors would like to acknowledge NSF CCI grant number: CHE-1037992.


\bibliographystyle{dcu}
\bibliography{QC4Chem}

\end{document}